\documentclass[a4paper,12pt]{article}

\usepackage{amsmath,amssymb,amsfonts}
\usepackage[dvips]{graphicx}
\usepackage{epsfig}

\usepackage{setspace}
\usepackage{color}
\usepackage{amsmath}
\usepackage{amsfonts}
\usepackage{verbatim}
\usepackage{amssymb}  
\usepackage{graphicx}
\usepackage{amsmath,amssymb,calc}
\usepackage{setspace}
\usepackage{color}
\usepackage{amsmath}
\usepackage{amsfonts}
\usepackage{verbatim}
\usepackage{amssymb}
\usepackage{graphicx}
\usepackage{array}
\usepackage{caption}
\usepackage{subcaption}
\usepackage{epstopdf}
\usepackage[pdf]{pstricks}
\usepackage{upgreek}
\usepackage{cmll}

\usepackage{latexsym}
\usepackage{braket}

\usepackage{cite}
\usepackage{amsmath,amssymb,amsfonts,graphicx}
\usepackage{epsfig}
\usepackage[left=2.4cm,top=3.3cm,right=2.4cm,bottom=3.3cm,bindingoffset=0cm]{geometry}

\newcommand{\beq}{\begin{eqnarray}}
\newcommand{\eeq}{\end{eqnarray}}
\newcommand{\bea}{\begin{eqnarray}}
\newcommand{\eea}{\end{eqnarray}}
\newcommand{\be}{\begin{equation}}
\newcommand{\ee}{\end{equation}}

\def\1{\mathbbm{1}}

\def\tA{{\cal A}}
\def\tF{{\cal F}}
\def\tr{\mathrm{tr}}
\def\t{\tilde}
\def\B{{\rm B}}

\numberwithin{equation}{section}

\begin{document}

\title{
\begin{flushright}\ \vskip -1.5cm {\small {IFUP-TH-2017}}\end{flushright}
\vskip 40pt
\bf{ \Large  A Solitonic Approach to Holographic Nuclear Physics}
\vskip 18pt}
\author{{\large Salvatore Baldino$^{(1)}$, Stefano Bolognesi$^{(1)}$,
    Sven Bjarke Gudnason$^{(2)}$}\\[0pt]{\large and Deniz Koksal$^{(1)}$} \\[20pt]
{\em \normalsize
$^{(1)}$Department of Physics ``E. Fermi'', University of Pisa
  and INFN Sezione di Pisa}\\[0pt]
{\em \normalsize
  Largo Pontecorvo, 3, Ed. C, 56127 Pisa, Italy}\\[3pt]
{\em \normalsize
$^{(2)}$Institute of Modern Physics, Chinese Academy of Sciences,
Lanzhou 730000, China
}\\[10pt]
{\normalsize emails:  salvatorebaldino@gmail.com, stefano.bolognesi@unipi.it, } \\
{\normalsize bjarke@impcas.ac.cn, d.koksal@studenti.unipi.it
}}
\vskip 10pt
\date{March 2017}
\maketitle
\vskip 0pt

\begin{abstract}

We discuss nuclear physics in the Sakai-Sugimoto model in the limit of  large number $N_c$ of colors and large 't Hooft coupling $\lambda$. 
In this limit the individual baryons are described by classical solitons whose size is much smaller than the typical distance at which they settle in a nuclear bound state.
We can thus use the linear approximation outside the instanton cores to compute the interaction potential. 
We find the classical geometry of nuclear bound states for baryon number up to eight. 
One of the interesting features that we find is that holographic nuclear physics provides a natural description for lightly bound states when $\lambda$ is large.
For the case of two nuclei, we also find the topology and metric of the manifold of zero modes and, quantizing it, we find that the ground state can be identified with the deuteron state. We discuss the relations with other methods in the literature used to study Skyrmions and holographic nuclear physics. 
We discuss $1/{N_c}$ and $1/\lambda$ corrections and the challenges to overcome to reach the phenomenological values to fit with real QCD.

\end{abstract}

\newpage

\tableofcontents

\section{Introduction}

The Sakai-Sugimoto (SS) model is a holographic dual model of QCD \cite{Sakai:2004cn,Sakai:2005yt}.  
It is a top down approach and consequently has very few parameters to fit.
Flavor dynamics are encoded in the low energy action for the gauge field on the flavor branes,
and the baryons of QCD are the instantonic configurations of that gauge theory \cite{Hong:2007kx,Hata:2007mb,Hashimoto:2008zw,Bolognesi:2013nja}.
Quantization of the degrees of freedom for an instantonic field of charge one creates a quantum system of states, whose transformation properties and quantum numbers are just right to interpret them as nucleons. 
Nuclear physics at low energy is thus turned into a multi-instanton problem in a curved five-dimensional background; this is the problem we discuss in the present paper. We will approach the problem of nuclei in the SS model from a ``solitonic perspective", in a way somehow different, or complementary to other approaches which already exist in the literature \cite{Hashimoto:2009ys,Kim:2009sr,Kim:2008iy,Kim:2011ey,Pahlavani:2010zzb,Pahlavani:2014dma,Hashimoto:2010je,Bolognesi:2013jba}.
We shall use many techniques developed in the context of nuclei within the Skyrme model, for example \cite{Adkins:1983ya,Adkins:1983nw,Jackson:1985bn,Braaten:1988cc,Verbaarschot:1986qi,Piette:1994ug,Leese:1994hb,Manko:2007pr,Irwin:1998bs}.

The limit which we consider is that of the large number of colors  $N_c$ and large 't Hooft coupling $\lambda$. The instanton radius scales as $\lambda^{-1/2}$ and, as we shall verify a posteriori, the distances between individual nuclei in the bound state configuration scale as $\lambda^0$. This suggests us to use a linear approach for the computation of the dominant two-body potential between the nuclei. Our first result is that nuclear physics at large  $N_c$ and large  $\lambda$ does have bound states in the linear regime.
In this picture, we build a charge two field configuration by ``gluing'' together two single charge instanton solutions, where by gluing we mean taking the linear superposition. 
In the large $N_c$ and $\lambda$ limit  we compute exactly the energy of this field configuration and interpret the result as the potential of interaction between nuclei.
This is proposed as a classical potential for the baryon interaction, where its structure as an infinite sum of Yukawa monopole and dipole interactions is interpreted as the classical analogue of the exchange interaction with a meson mediator.
We show how classical nuclei with multiple baryons can be described in this limit.
The solution has some analogies with the one obtained recently in a lightly bound Skyrme model \cite{Gillard:2015eia,Gillard:2016esy}.
We confront our potential with the one obtained in \cite{Kim:2009sr} through a different approach and explain the differences and the limits of validity for the various approaches at hand.

Focusing then on the two-nuclei system, we quantize the coordinates of the two instanton fields and impose physical constraints in order to restrict the spectrum of the system.
We see that the internal degrees of freedom of the system can be rearranged and interpreted as the total spin and the isospin of the system, and that they assume only integer values. 
Among the states that are compatible with our constraints, we find one with the right angular quantum numbers (spin one and isospin zero) to be interpreted as the deuteron. 

In  section \ref{due}, we review the low energy action of the SS model, concentrating on the solitonic solutions of the theory.  
In section \ref{tre},  by gluing together two solutions at large spatial separation, we find a classical interaction potential between the nucleons. We then  generalize to topological sectors of arbitrarily high charge.
In section \ref{quattro} the $\B=2$ system is quantized and we show that the minimal energy state in the spectrum has the same quantum numbers as the deuteron.
In section \ref{cinque} we discuss various types of corrections from the inclusion of the massive modes.
We conclude in section \ref{sei}.

\section{Holographic baryons in the Sakai-Sugimoto model}
\label{due}

We take, as a starting point, the low-energy action of the SS model, which is a five dimensional gauge theory in a particularly curved background.
We call the gauge field $\tA$, and its associated field strength tensor $\tF$.  In these terms, we are studying a field theory of the form
\bea
\tA:\mathcal R \longrightarrow u(N_f),
\eea
where the space-time $\mathcal R$ has the topology of $\mathbb R^{4|1}$, and the metric is given by
\bea
g=H(z)\eta_{\mu\nu}dx^\mu dx^\nu+\frac{1}{H(z)}dz^2,
\label{metricm}
\eea
and the warp factor, $H(z)$, is
\bea
H(z)=\left(1+ M_{\rm KK}^2z^2\right)^{\frac 2 3}.
\eea
From now on, we adopt the units of $M_{\rm KK}=1$. The action is given by
\bea
 S&=&-\frac{N_c \lambda}{216\pi^3}\int\sqrt{|g|}\frac{1}{2}\tr\left(\tF_{\Gamma\Delta}\tF^{\Gamma\Delta}\right)d^4xdz+  \frac{N_c}{24\pi^2}\int\omega_5(\tA)d^4x dz.
\label{main}
\eea
The term in the second integral is the Chern-Simons term. 
$N_c$ is the number of colors from the dual QCD and it is an overall multiplicative constant of the above action. 
The classical equations of motion are thus completely independent of $N_c$ and the quantum corrections are negligible when we take the $N_c \to \infty$ limit.

We divide the field into two components: an abelian $\hat A$ and a non-abelian part $A$. Similarly for the field strength
\bea
\tA_\Gamma= A_\Gamma+\frac{1}{N_f}\hat A_\Gamma,\quad\quad\tF_{\Gamma\Xi}=F_{\Gamma\Xi}+\frac{1}{N_f}\hat F_{\Gamma\Xi}.
\eea
We rescale the action as
\bea
\mathcal S=\frac{216\pi^3}{N_c\lambda}S,
\eea
and define the new coupling
\bea
\Lambda=\frac{8\lambda}{27\pi}.
\eea
Furthermore, we restrict to the case of two flavors $N_f=2$ for simplicity. 
The rescaled action reads
\bea
\nonumber
\mathcal{S} &=& -\frac12\int H^{\frac32}(z) \left(\frac12\hat{F}_{\Gamma\Sigma}\hat{F}^{\Gamma\Sigma}+\tr(F_{\Gamma\Sigma}F^{\Gamma\Sigma})\right)d^4xdz\\
&&\mathop+\frac{1}{\Lambda}\int\left(\hat A_\Gamma\tr(F_{\Delta\Sigma}F_{\Xi\Phi})+\frac{1}{6}\hat A_{\Gamma}\hat F_{\Delta\Sigma}\hat F_{\Xi\Phi}\right)\epsilon^{\Gamma\Delta\Sigma\Xi\Phi}d^4xdz.
\eea

\subsection{Classical baryon solution}

We want to find static solutions of this theory. To do so, we perform the static ansatz as
\bea
A_{I}=A_{I}(x_{J}),\quad\quad A_0=0,\quad\quad\hat A_{I}=0,\quad\quad \hat A_0=\hat A_0(x_I),
\eea
that is, we remove all dependence of time coordinates from the fields $\hat A_I$ and the field $A_0$.  In this ansatz, we also suppose that $\hat A_0$ is not a propagating field, but a constrained field fixed by the equations of motion.  With this ansatz, the action reads
\bea
\nonumber
\mathcal S &=& \int\left(\frac{1}{2H^{\frac 1 2}}(\partial_i\hat A_0)^2+\frac{H^{\frac 3 2}}{2}(\partial_z \hat A_0)^2-\frac{1}{2H^{\frac 1 2}}\tr(F_{ij}^2)-H^{\frac 3 2}\tr(F_{iz}^2)\right)d^4x dz\\
&& \mathop+\frac{1}{\Lambda}\int\hat A_{0}\tr(F_{IJ}F_{KL})\epsilon_{IJKL}d^4x dz,
\label{staticaction}
\eea
and the equations of motion are
\bea
&&\frac{1}{H^{\frac 1 2}}D_{j}F_{ji}+D_z(H^{\frac 3 2}F_{zi})=\frac{1}{\Lambda}\epsilon_{iJKL}F_{KL}\partial_J \hat A_0,\label{motiona}\\
&&H^{\frac{3}{2}}D_j F_{jz}=\frac{1}{\Lambda}\epsilon_{ijk}F_{jk}\partial_{i}\hat A_0, \label{motionb}\\
&&\frac{1}{H^{\frac{1}{2}}}\partial_i\partial_i\hat A_0+\partial_z(H^{\frac{3}{2}}\partial_z\hat A_0)=\frac{1}{\Lambda}\tr(F_{IJ}F_{KL})\epsilon_{IJKL},\label{lastz}
\eea
where $D$ is the covariant derivative with respect to the field $A$.
The last equation defines $\hat A_{0}$ as the inhomogeneous solution of the equation, obtained through convoluting the Green function of the LHS operator with the RHS.

To have a finite action solution, the non-abelian gauge field must
approach a pure gauge configuration on the sphere at infinity, $S^{3}_{\infty}$:
\bea
A_{I}(x^I)|_{S^{3}_{\infty}}=g^\dagger\partial_Ig, \quad\quad g:S^3_{\infty}\to SU(2).
\eea
As $\pi_{3}(SU(2))=\mathbb Z$, we have a discrete (but infinite) number of topological sectors, labeled by the topological charge
\bea
\B=\int B^0(x,z)d^3xdz=-\frac{1}{32\pi^2}\int\tr(F_{IJ}F_{KL})\epsilon_{IJKL}d^3x dz,
\label{chargeden}
\eea
that assumes integer values. We have an additional constrained field, $\hat A_0$, that can be interpreted as an electrostatic potential for the electric field $\hat F_{0I}=-\partial_I \hat A_0$, sourced by the topological charge.

We review the solution for the $\B=1$ sector
\cite{Hata:2007mb,Bolognesi:2013nja}. We assume a central ansatz
$A_I=A_I(\rho)$, with $\rho=\sqrt{x^Ix^I}$, even if the curvature
along the $z$ direction explicitly breaks invariance with respect to translations along $z$. We make the 't Hooft ansatz
\bea 
A_I=-\sigma_{IJ}\partial_{J}b(\rho),
\quad\quad
\hat A_0=a(\rho),	\label{ansatz}
\eea
where 
\bea
\sigma_{ij}=\epsilon_{ijk}\sigma_{k},\quad\quad\sigma_{zi}=\sigma_i,\quad\quad \sigma_{IJ}=-\sigma_{JI}.
\eea
The appropriate boundary conditions to have finite energy and $\B=1$ are
\bea
\lim_{\rho\to\infty}\rho^2b(\rho)=1,\quad\quad b'(0)=0.
\eea
 Inserting the ansatz in the action and developing in orders of $1/\Lambda$ we see that, at order $\Lambda^0$ in the scaled action and neglecting warp factors, we have the same action of the BPST instanton:
\bea
b(\rho)=\frac{1}{\Lambda(\rho^2+\mu^2)},
\eea
where $\mu^2$ represents the instanton size, which is a modulus for the standard BPS instanton. The rescaled energy $\mathcal E=-\mathcal S$ is given by
\bea
\mathcal E=2\pi^2\left(4+\frac 2 3\mu^2+\frac{256}{5\Lambda^2\mu^2}\right)+O(\Lambda^{-3}).
\label{restene}
\eea
 The energy of the instanton then grows with its size, and with the gravitational effect alone the instanton becomes pointlike and placed at $z=0$.  The instanton would shrink to zero size, would it not be for the Chern-Simons term: the abelian field $\hat A_0$ acts as an effective electric potential, and as the topological charge density is positive everywhere the net effect of the electric field is to expand the instanton. Those two effects combine to give an instanton of definite classical size
\bea
\mu=\frac{4}{\sqrt\Lambda}\left(\frac 3 {10}\right)^{\frac 1 4}.\label{size}
\eea
As the energy is size dependent, $\mu$ is not a modulus for the SS
instanton, and it is fixed to the value \eqref{size} unless stated otherwise. $a$ is given by
\bea
a(\rho)=\frac{8}{\Lambda}\frac{\rho^2+2\mu^2}{(\rho^2+\mu^2)^2}.
\eea
In normal units, the soliton has energy (that we interpret as rest mass)
\bea
E=M=\frac{N_c\Lambda}{8}+\sqrt{\frac{2}{15}}N_c.
\eea
The presence of a gauge field used to stabilize a soliton is not a peculiarity of this model, and it has been amply studied as an alternative term used to stabilize the Skyrmion, see for example \cite{Adkins:1983nw,Sutcliffe:2008sk}.

We now turn our attention to the moduli space of zero modes. We have explicit translational invariance along the $x^i$ coordinates, so we have three moduli $X^i$, indicating the position of the instanton in physical space. We also have global gauge transformations, which do not fall off to zero at infinity. We get as the moduli space
\bea
\mathcal M=\mathbb R ^3\times (SU(2)/\mathbb{Z}_2).
\eea
The calculation of the metric on the moduli space is similar to the standard calculation for the standard BPS instanton, and the result is the same \cite{Hata:2007mb}. It reads
\bea
g|_{\mathcal M}=dx^i dx^i+2\mu^2 d\Omega_{SU(2)},
\label{metsing}
\eea
where $d\Omega_{SU(2)}$ is the standard $SU(2)$ metric. 
The instanton size $\mu$ and the coordinate along the $z$ direction are massive moduli.

\subsection{The linear regime}

We now perform an expansion in $1/\Lambda$. The objective is to find an expression for the fields and the equations of motion \eqref{motiona}, \eqref{motionb} and \eqref{lastz} and identify the \textit{linear region} of the soliton, the region of space where we can approximate the gauge potential with its first term in the $1/\Lambda$ expansion \cite{Hashimoto:2008zw,Bolognesi:2013nja}.

We define the $1/\Lambda$ approximation through
\bea
A_{I}=A_{I}^{(1)}+A_{I}^{(2)}+ \dots
\eea
where each term $A_{I}^{(n)}$ is of order $1/\Lambda^n$. We are
interested in the equations of motion for the field $A_{I}^{(1)}$. 
In the linear zone (which is given by $\rho>1/\sqrt\Lambda$), we can take only the $A_{I}^{(1)}$ contributions to the action and the equations of motion, effectively linearizing the system.\footnote{Actually, the linear approximation is valid up to $\rho<\ln\Lambda$: in the region $\rho>\ln\Lambda$ the contributions $A_{I}^{(n)}$ with $n>1$ become more important than $A_I^{(1)}$, so the linear approximation breaks down in that region \cite{Bolognesi:2013nja}. We will consider the situation where $\Lambda \gg1$ and neglect that zone.}
Before proceeding, we divide the field 
\beq
A_i=A_i^{+}+A_i^{-} \ ,
\eeq 
where the superscript indicates parity with respect to $z\to-z$: the
$A_z$ part is an even function, in the gauge where the core potential
has been obtained, so $A_z=A_z^+$. Restricting to the order
$1/\Lambda$ terms in the equations of motion (and dropping the $(1)$ superscript), we have

\bea
\frac{\partial_{i}\partial_{i}}{H^{\frac{1}{2}}}\hat{A}_0+\partial_{z}(H^{\frac{3}{2}}\partial_{z}\hat{A}_0) &=&  \text{source1}\ ,\\
\frac{\partial_{j}\partial_{j}}{H^{\frac{1}{2}}}A_{i}^{+}+\partial_{z}(H^{\frac{3}{2}}\partial_{z}A_{i}^{+}) &=& \text{source2}\ ,\\
H^{\frac{3}{2}}(\partial_{i}\partial_{i}A_{z}^{+}-\partial_{i}\partial_{z}A_{i}^{-}) &=& \text{source3}\ ,\phantom{\frac12}\\
\frac{\partial_{j}\partial_{j}A_{i}^{-}-\partial_{j}\partial_{i}A_{j}^{-}}{H^{\frac{1}{2}}}-\partial_{z}(H^{\frac{3}{2}}(\partial_{i}A_{z}^{+}-\partial_{z}A_{i}^{-})) &=&\text{source4}\ ,
\eea
where the source terms are delta functions or derivatives, centered at $(x,z)=(0,0)$. By developing the core solution to first order in $1/\Lambda$, we obtain explicit expressions for $\hat A_0, A_I$ and use them to calculate the source terms.

To proceed, we define the  $\psi_n$ functions to solve the equation
\begin{align}
H^{\frac 1 2}\partial_z(H^{\frac 3 2}\partial_z\psi_n(z))+k_n^2\psi_n(z)=0
\label{equno}
\end{align}
for some numbers $k_n$.
We define the scalar products
\begin{align}
(f,g)=\int_{-\infty}^{+\infty}\frac{f(z)g(z)}{H^{\frac{1}{2}}}dz
,\quad\quad<f,g>\mathop{=}\int_{-\infty}^{+\infty}H^{\frac 3 2}f(z)g(z) dz.
\end{align} This way, by partial integration, we can see that
\begin{align}
(\psi_n,\psi_m)  = \frac{<\psi'_n,\psi'_m>}{k_n^2},
\end{align}
we can just set $\phi_n=\psi'_n$. The $\phi_n$ obey the differential equation
\begin{align}
\partial_z(H^{\frac{1}{2}}\partial_z(H^{\frac 3 2}\phi_n(z)))+k_n^2\phi_n(z)=0.
\label{eqdue}
\end{align}

\begin{figure}[h]
\centering
\includegraphics[scale=0.9]{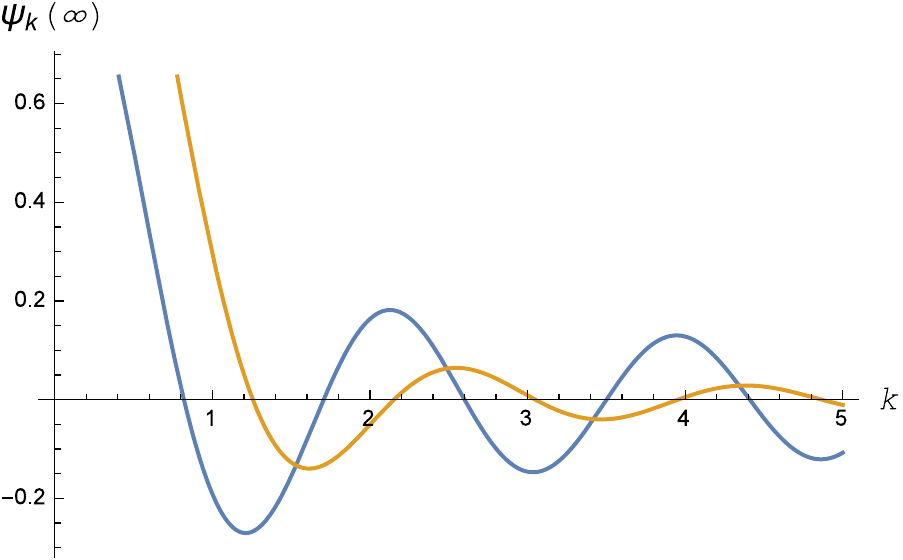}
\caption{Asymptotic value of $\psi_k(+\infty)$, plotted against $k$ for even and odd normalization. Their zeroes indicate a meson mass.}
\label{asymptotics}
\end{figure}

We can numerically calculate the functions $\psi_n$ in the following way. Let $f(k)$ be the asymptotic value for $z\to+\infty$ of an even solution to \eqref{equno} with $k$ in place of $k_n$, and let $g(k)$ be the same for odd functions. Searching for normalizable solutions of \eqref{equno} then amounts to finding the zeroes of $f(k)$ and $g(k)$. We plot those functions in Fig.~\ref{asymptotics}. The zeroes of $f(k)$ are the entries of $k_n$ with odd $n$, while the zeroes of $g(k)$ are the entries of $k_n$ with even $n$.

There is a subtlety: if we set $k_n=0$, we see that $\phi_n\propto
H^{-\frac 3 2}$ is a solution. We also have that $<\phi_0,\phi_0>$
converges, and it has the value $\pi$, so we can include $\phi_0$ in the expansion of $A_z$.
We note that the primitive of $\phi_0$, that would be $\psi_0 \propto \frac2\pi \arctan z$, still solves $\eqref{equno}$, but does not fall off at infinity and is not normalizable under the scalar product $(\psi_0,\psi_0)$.

\begin{figure}[h]
\centering
\includegraphics[scale=.75]{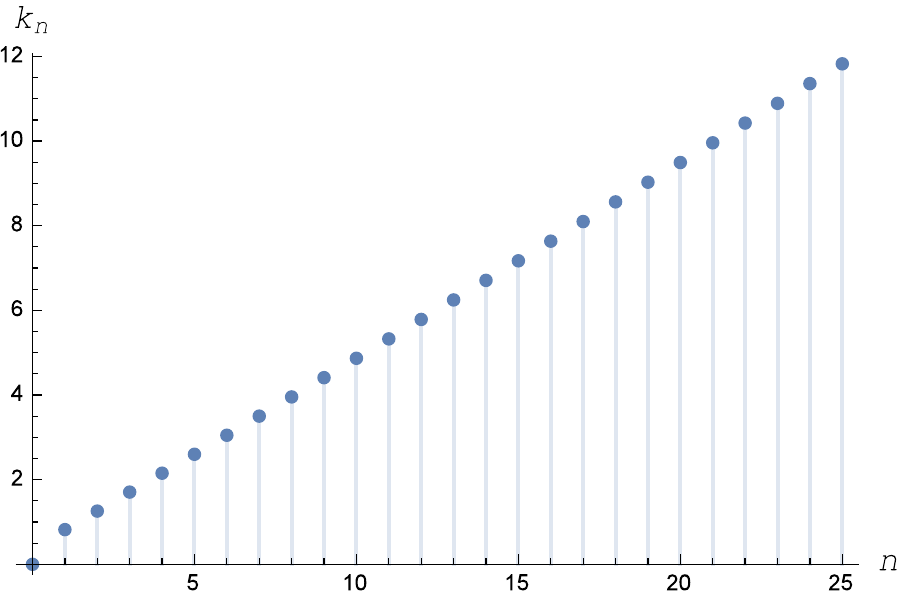} \qquad 
\includegraphics[scale=.75]{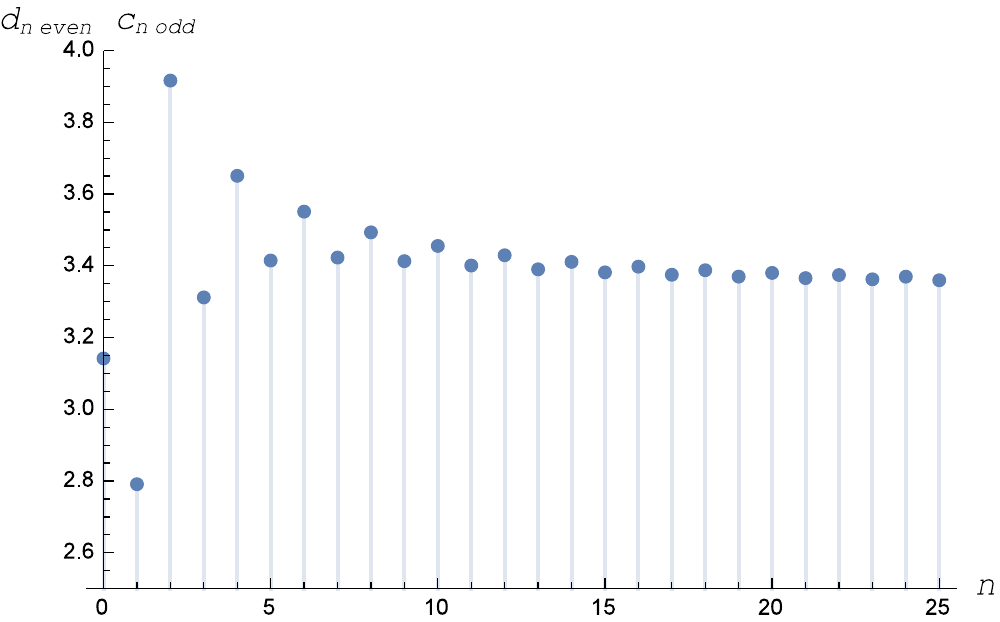}
\caption{On the left we plot the values of the pulses $k_n$; on the right we plot the values of $c_n$ when $n$ is odd, $d_n$ when $n$ is even.}
\label{norm}
\end{figure}

 We impose $(\psi_n(0)=1,\psi_n'(0)=0)$ for $n$ odd and $(\psi_n(0)=0,\psi_n'(0)=1)$ for $n$ even, where the prime is the derivative with respect to $z$. This way, we have that $\psi_n(-z)=(-1)^{n+1}\psi_n(z)$. We define
\bea
(\psi_n,\psi_m)=c_n\delta_{nm},\quad\quad<\phi_n,\phi_m>\mathop=d_n\delta_{nm}.
\eea
where $c_n$ and $d_n$ have to be determined numerically. As $k_n^2(\psi_n,\psi_m)\mathop{=}<\psi'_n,\psi'_m>$ we have $k_n^2c_n=d_n$. The only particular value is the norm of $\phi_0(z)=H^{-\frac32}(z)$: we have $d_0=\pi$, while $c_0$ is divergent.  In the potential we will have to use as coefficients the $c_n$ with $n$ odd and the $d_n$ with $n$ even. We plot the values of the pulses $k_n$ and the alternating succession of $c_n$ and $d_n$ in Figure \ref{norm}.

With the previous choice of  normalization, the completeness relations are
\bea
\sum_{n=1}^{\infty}\frac{\psi_n(z)\psi_n(z')}{H^{\frac 1 2}(z)c_n}=\delta(z-z'),\quad\quad\sum_{n=1}^{\infty}H^{\frac 32}(z)\frac{\phi_n(z)\phi_n(z')}{d_n}=\delta(z-z').
\eea
We thus define, following \cite{Hashimoto:2008zw}, the Green functions
\bea
 G(x,z,x',z')&=&-\frac{1}{4\pi}\sum_{n=1}^{\infty}\frac{\psi_n(z)\psi_n(z')}{c_n}\frac{e^{-k_n|x-x'|}}{|x-x'|}\ , \nonumber \\
 L(x,z,x',z')&=&-\frac{1}{4\pi}\sum_{n=0}^{\infty}\frac{\phi_n(z)\phi_n(z')}{d_n}\frac{e^{-k_n|x-x'|}}{|x-x'|}\ ,
\eea
which obey
\bea
\frac{\partial_i\partial_i G}{H^{\frac12}(z)}+\partial_z(H^{\frac32}(z)\partial_z G)&=&\delta^3(x-x')\delta(z-z')\ ,\nonumber\\
\partial_i\partial_i L-\partial_z\partial_{z'}G &=&\delta^3(x-x')\delta(z-z')\ , \phantom{\frac12}\nonumber\\
\partial_z(H^{\frac32}(z)L)+H^{-\frac12}(z)\partial_{z'}G &=& 0 \phantom{\frac12}.
\eea

We now take the linear approximation to the core solution from \cite{Bolognesi:2013nja}. In terms of the functions $G$ and $L$, they can be written as
\bea
\hat A_0(x,z)&=&-\frac{32\pi^2}{\Lambda}G(x,z,0,0)\ , \nonumber\\
A_i^{+}(x,z)&=&\mathop-2\pi^2\mu^2\epsilon_{ijk}\sigma_k\partial_jG(x,z,0,0)\ , \phantom{\frac12}\nonumber\\
A_i^-(x,z)&=&-2\pi^2\mu^2\sigma_i\partial_{z'}G(x,z,0,z')|_{z'=0}\ ,  \phantom{\frac12}\nonumber\\  
A_z^+(x,z)&=&-2\pi^2\mu^2\sigma_i\partial_iL(x,z,0,0) \phantom{\frac12}.
\label{sol1}
\eea
We now apply the operators of the linear equations of motion, obtaining the form of the source terms:
\bea
 \frac{\partial_{i}\partial_{i}}{H^{\frac{1}{2}}}\hat{A}_0+\partial_{z}(H^{\frac{3}{2}}\partial_{z}\hat{A}_0) &=&-\frac{32\pi^2}{\Lambda}\delta^3(x)\delta(z) \ ,\label{motlinmon}\nonumber\\
 \frac{\partial_{j}\partial_{j}}{H^{\frac{1}{2}}}A_{i}^{+}+\partial_{z}(H^{\frac{3}{2}}\partial_{z}A_{i}^{+}) &=&-2\pi^2\mu^2\epsilon_{ijk}\sigma_k\partial_j\delta^3(x)\delta(z)\ ,\label{motlina}\nonumber\\
 H^{\frac{3}{2}}(\partial_{i}\partial_{i}A_{z}^{+}-\partial_{i}\partial_{z}A_{i}^{-}) &=&-2\pi^2\mu^2\sigma_i\partial_i\delta^3(x)\delta(z)\ ,\label{motlinb}\nonumber\\
 \frac{\partial_{j}\partial_{j}A_{i}^{-}-\partial_{j}\partial_{i}A_{j}^{-}}{H^{\frac{1}{2}}}-\partial_{z}(H^{\frac{3}{2}}(\partial_{i}A_{z}^{+}-\partial_{z}A_{i}^{-})) &=& 2\pi^2\mu^2\sigma_i\delta^3(x)\partial_{z}\delta(z)\ .\label{motlinc}
\eea

We can generalize the linear form with an arbitrary $SU(2)$ phase $G$ and an arbitrary $\mathbb R^3$ position, $X$: this is done by substituting $G(x,z,0,0)$ with $G(x,z,X,0)$ (and analogously for $L$), and every occurrence of the Pauli matrices $\sigma_i$ with $G\sigma_iG^\dagger$.
We will use this linear form of the fields in the following when calculating the interaction potential between nuclei.

\section{Nucleon-Nucleon potential and  classical nuclei}
\label{tre}

\begin{figure}[h]
\centering
\includegraphics[scale=1.3]{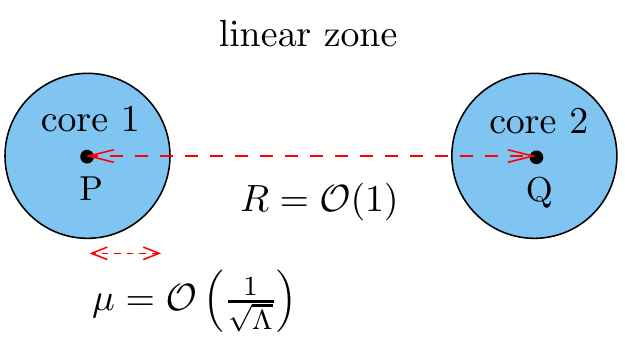}
\caption{Soliton configuration for the charge two sector.}
\label{conf}
\end{figure}

We now perform the calculation of the holographic potential between nucleons. 
To do so, we place the instantons with their cores at a distance $R$ from each other, which we assume to be greater than $\sim \mu$ and we set both holographic coordinates for the two instantons to zero in order to minimize the energy.
The system is diagrammatically shown in Figure \ref{conf}.  We write the single instanton fields by writing the first one as in \eqref{sol1} and writing the second one by translating it to $(R,0,0)$ and assigning an arbitrary phase matrix $G$ to it.
We call $\tA^p$, the gauge field centered in the origin, $(0,0,0)$, and $\tA^q$, the gauge field centered in $(R,0,0)$. Due to the distance between the fields, we can take the gauge field in the whole space to be $\tA^p+\tA^q$: in the ``core 1'' region, $\tA^q$ is small and can be considered as a small perturbation, while the opposite situation happens in  ``core 2''. There is a linear zone where both fields are weak, and can be both approximated by their linear form.

\subsection{The interaction potential}

The energy of the configuration can be found by using the fact that the $\B=2$ field can be approximated by the sum of two $\B=1$ fields and one of the coefficients of the sum can always be taken as a linear perturbation. We start by writing the scaled energy, through an integration by parts:
\bea
\mathcal E=\int\left(\frac{1}{2H^{\frac{1}{2}}}\tr(F_{ij}^2)+H^{\frac 3 2}\tr(F_{iz}^2)-\frac{1}{2}\hat A_0\left(\frac{\partial_i\partial_i}{H^{\frac 1 2}}+\partial_z(H^{\frac 3 2}\partial_z)\right)\hat A_0\right)d^3xdz.
\label{entwo}
\eea
In the integration by parts, we have used the fact that the functions $\hat A_0$ are supposed to vanish at the boundaries fast enough for the energy to be finite. We split this integral into two: we will see that the first two terms (called $\mathcal E_1$) give the dipole interaction contribution, while the last one (called $\mathcal E_2$) gives a monopole interaction.

Let us start with the evaluation of the monopole term
\bea
\mathcal E_{2}=-\int\frac{1}{2}\hat A_0\triangle \hat A_0d^3xdz \ ,
\label{monopole}
\eea
where $\triangle$ is the Laplace-Beltrami operator
\beq
\triangle = \frac{\partial_i\partial_i}{H^{\frac 1 2}}+\partial_z(H^{\frac 3 2}\partial_z) \ .
\eeq
In our approximation, we can divide the topological charge density as $B^0\simeq B^{0,p}+B^{0,q}$, so that we also divide the gauge field into $\hat A^p+\hat A^q$, such as $\triangle \hat A^p=-(32\pi^2\Lambda^{-1})B^{0,p}$, and similarly for $\hat A^q$. The monopole term \eqref{monopole} becomes
\bea
\mathcal E_{2}=-\frac{1}{2}\int \left(\hat A_0^p\triangle \hat A_0^p+\hat A_0^q\triangle \hat A_0^q+\hat A_0^p\triangle \hat A_0^q+\hat A_0^q\triangle \hat A_0^p\right)\,d^3xdz\ .
\label{monoevaluate}
\eea
The terms $\hat A_0^p\triangle \hat A_0^p+\hat A_0^q\triangle \hat
A_0^q$ contribute to the self energies of the instanton and we neglect
them as we are really interested in the cross terms in order to obtain
the  potential. Let us then take $\hat A_0^p\triangle\hat
A_0^q=-32\pi^2\Lambda^{-1}\hat A^{p}B^{0,q}$. $B^{0,q}$ is peaked in
the $q$ zone, where $\hat A^p$ must be taken as its linear
approximation. We can then suppose $B^{0,q}$ to be strongly localized
at $(R,0,0)$ through the delta function:
$B^{0,q}\simeq\delta^{3}(x-R)\delta(z)$. In this approximation, the
topological charge of the soliton $\tA_q$ is still one. Any
contribution that tends to enlarge the soliton comes from the
electrostatic field, and is then multiplied by some negative power of
$\Lambda$: as we are keeping the linear order in $\Lambda$ we can neglect those contributions. With the $\delta$ functions, the integral is easily performed and we can do the same with the other term too. Summing everything and removing the self energies, we obtain the monopole part of the potential. Using the linear form of the fields, we have
\bea
{\cal V}_m=\frac{16\pi^{2}}{\Lambda}\left(\hat A^p_0(R,0)+\hat A^q_{0}(0,0)\right)=\frac{256\pi^3}{\Lambda^2}\sum_{n=1}^{\infty}\frac{1}{c_{2n-1}}\frac{e^{-k_{2n-1}R}}{R}.
\eea
This is the monopole potential, where only the contribution of $k_n$ with odd $n$ matters. This monopole interaction can be interpreted as a classical analogue of the exchange potential between the instantons, which interact by exchanging mesons $\omega_{2n-1}$ with masses $k_{2n-1}$.

The contribution of the dipole part can be calculated through a trick, similar to the one used in \cite{Piette:1994ug}. Dividing the space into $P$ (core 1), $Q$ (core 2) and $LZ$ (linear zone), we split the integral as
\bea
\int_{\mathcal R}=\int_P+\int_Q +\int_{LZ}\ .
\eea
In the $P$ region, we can take, as a first approximation, the whole gauge field to be coincident with $A_I^p$. Then, we relax this approximation by admitting variations of the form $\delta A_I^p=A_I^q$, always taking the first order in $A_I^q$. The integral over the $P$ region of the unperturbed field is a contribution to its self energy, while the variation of this energy accounts for the interaction between the instantons and consequently is the only piece that we need. The variations that we need are:
\bea
\delta F_{IJ}^p=D^p_{I}A_J^q-D^p_JA_I^q,\quad\quad\delta\int\tr(F_{IJ}^pF_{IJ}^p)d^3xdz=4\int\tr(F^p_{IJ}D_I^p A^q_J)d^3xdz \ ,
\eea
where we denote the field strength and the covariant derivative built from $A_I^p$ as  $F^p$ and $D^p$ respectively. We can do the same in the $Q$ and $LZ$ region, interchanging the roles of the two fields.  Noting
\bea
P_{ij}^{(p,q)}=\frac{2 F_{ij}^{(p,q)}}{H^{\frac 1 2}},\quad\quad P_{iz}^{(p,q)}=2H^{\frac 3 2}F_{iz}^{(p,q)}\ ,
\eea
we can write
\bea
{\cal V}_{d}=\int_{P}\tr(P^p_{IJ}D_I^pA_J^q)d^3xdz+\int_{Q\cup LZ}\tr(P^q_{IJ}D_I^qA_J^p)d^3xdz\ .
\eea
Since the gauge field in the core region goes as $1/\Lambda$ for great $\Lambda$ and so does the linear approximation, we can approximate the covariant derivative with the usual one. We can then use Stokes's theorem, using the fact that $\partial P=-\partial (Q\cup LZ)$, to get
\bea
{\cal V}_{d}=\int_{\partial P}(P^p_{IJ}A_J^q-P^q_{IJ}A_J^p)\,d\Gamma_I\ ,
\eea
where $d\Gamma_I$ is a vector field normal to $\partial P$, pointing
outwards (remember that $P$ is a ball in four dimensions). In the
region $\partial P$, both fields take their linear form, so we can
linearize the field strength tensors (neglecting the commutator) and
approximate every $A^{(p,q)}$ with their linear approximations. We use
Stokes' again to return inside the $P$ region. Derivatives act only on
the field strength, as when they act on the gauge field, the first
term cancels the second one. Using the linear equations of motion, we
have $\partial_IP_{IJ}^q=0$, as we are integrating in the $P$ region
and the core of $A^q$ is outside of it. Performing the division in parity components, we get the integral
\bea
{\cal V}_{d} &=& 2\int_{P}\tr\left(A_{i}^{+,q}\left(\frac{\partial_{j}\partial_{j}A_{i}^{+,p}}{H^{\frac{1}{2}}}+\partial_{z}(H^{\frac{3}{2}}\partial_{z}A_{i}^{+,p})\right)\right)d^3xdz\label{dipole}\\\nonumber
 && \mathop+2\int_P H^{\frac{3}{2}}\tr\left(A_{z}^{+,q}\left(\partial_{i}\partial_{i}A_{z}^{+,p}-\partial_{i}\partial_{z}A_{i}^{-,p}\right)\right)d^{3}xdz \phantom{\left(\frac12\right)} \nonumber \\
&&\mathop+2\int_{P}{\tr\left(A_{i}^{-,q}\left(\frac{\partial_{j}\partial_{j}A_{i}^{-,p}-\partial_{j}\partial_{i}A_{j}^{-,p}}{H^{\frac{1}{2}}}-\partial_{z}(H^{\frac{3}{2}}(\partial_{i}A_{z}^{+,p}-\partial_{z}A_{i}^{-,p}))\right)\right)d^{3}xdz}\ . \nonumber 
\eea
Using the equations of motion \eqref{motlina}, we see that the operators in the parentheses, when applied to the $A^p$ fields, give terms proportional to a Dirac delta function, such that the integrals are simply done by evaluating $A^q$ at the origin and then adding the necessary constants and derivatives.

The first line of the potential reads
\bea
{\cal V}_{d,1}=\frac{256\pi^{3}}{\Lambda^{2}}\frac{6}{5}\sum_{n=1}^{\infty}{\frac{1}{c_{2n-1}}\left(M_{ii}k_{2n-1}^{2}R^{2}\frac{e^{-k_{2n-1}R}}{R^{3}}-M_{ij}\partial_{i}\partial_{j}\frac{e^{-k_{2n-1}R}}{R}\right)}\ .
\eea
Here we have used the explicit form of $\mu^2$ \eqref{size} in order to obtain the $\Lambda^{-2}$ dependence, just as we did with the monopole term. The matrix $M_{ij}=M_{ij}(G)$ is equal to
\bea
M_{ij}(G)=\frac{1}{2}\tr\left(\sigma_i G\sigma_j G^\dagger\right).
\label{associated}
\eea
This term can be interpreted as the sum of Yukawa dipole interactions between the two instantons, mediated by the infinite tower of mesons, $\rho_{2n-1}$, which have the same masses as the $\omega_{2n-1}$ mesons. While the monopole interaction is always repulsive, the dipole interaction depends on the phase matrix $G$, which is interpreted as the isorotation that we must perform on the first object in order to obtain the same iso-orientation from the second object.

The last part of the potential comes from the last two lines of \eqref{dipole}. They are combined in the term
\bea
{\cal V}_{d,2}=-\frac{256\pi^{3}}{\Lambda^{2}}\frac{6}{5}\sum_{n=0}^{\infty}{\frac{1}{d_{2n}}\left(M_{ii}k_{2n}^{2}R^{2}\frac{e^{-k_{2n}R}}{R^{3}}-M_{ij}\partial_{i}\partial_{j}\frac{e^{-k_{2n}R}}{R}\right)}.
\eea
There are some fundamental differences between $\mathcal V_{d,1}$ and $\mathcal V_{d,2}$. The first one is the overall sign. The second is  the fact that we are also including a $k_0$ contribution: as $k_0=0$, $\mathcal V_{d,2}$ contains a massless, long range interaction. The particle that we classically take as the mediator of this long range interaction is the pion, which is massless in our model. The other mesons, of mass $k_{2n}$, are interpreted as a tower of $a_{2n}$ mesons.

Now that we have a final result for the interaction potential, we scale back to physical units and perform some changes in order to have a more generalized result which we will use in the following sections. We denote the coordinates of the first instanton by $(X_1,B)$, and the coordinates of the second instanton by $(X_2,C)$, where $X_n$ are 3-vectors and $B$ and $C$ are $SU(2)$ matrices. The field configuration described by this coordinate configuration is
\bea
BA_I(x-X_1)B^\dagger+CA_I(x-X_2)C^\dagger.
\eea
We make the change of variables: $X_{1,i}-X_{2,i}=r_i$ and
$R_i=(X_{1,i}+X_{2,i})/2$, as is usually done in two-body problems:
hence the potential will only depend on the relative distance $r_i$. It is also easy to find out that $G$ has to be substituted simply by $B^{\dagger}C$, indicating the relative orientation of the two objects. We define the symmetric tensor
\bea
P_{ij}(r,k)=\delta_{ij}((r k)^2+rk+1)-\frac{r_i r_j}{r^2}((rk)^2+3rk+3),
\eea
with $r$ on the RHS indicating the modulus of the position vector, and express the potential as
\bea
\nonumber V(r,B^{\dagger}C)&=&\frac{4\pi N_c}{\Lambda}\left(\sum_{n=1}^{\infty}\left(\frac{1}{c_{2n-1}}\frac{e^{-k_{2n-1}r}}{r}+\frac{6}{5}\frac{1}{c_{2n-1}}M_{ij}(B^{\dagger}C)P_{ij}(r,k_{2n-1})\frac{e^{-k_{2n-1}r}}{r^{3}}\right.\right.\\ 
&& \ \  \ \ \nonumber \left.\left.\mathop-\frac{6}{5}\frac{1}{d_{2n}}\frac{e^{-k_{2n}r}}{r^{3}}M_{ij}(B^{\dagger}C)P_{ij}(r,k_{2n})\right)\right.\\&& \ \  \ \ \ -\left.\frac{6}{5\pi}\frac{1}{r^{3}}\left(\tr M-3\frac {r\cdot M\cdot r}{r^2}\right)\right) \ . 
\label{potential}
\eea
We have separated the pion contribution from the rest of the $a$ meson tower and used explicitly $d_0=\pi$ and $P_{ij}(r,0)=\delta_{ij}-3(\frac{r_ir_j}{r^2})$.

\subsection{Looking for a bound state: the classical deuteron}

We can obtain a classical description of the deuteron by looking for a minimum energy configuration, where we choose the coordinates of our instantons to minimize \eqref{potential}.

We have to choose the relative orientation of the instantons. To do that, it is useful to switch to the axis-angle notation in order to write the rotation matrix $M_{ij}$. As $M_{ij}$ is an $SO(3)$ matrix, it can be specified by giving two components of a versor, the axis of rotation $n$ (where the third component is decided from the normalization of the vector, with a positive sign), and an angle $\alpha$, indicating the rotation around the versor (counterclockwise). We can then express any $M$ as
\bea
M_{ij}=\delta_{ij} \cos\alpha+(1-\cos\alpha)n_in_j+\epsilon_{ijk}n_k\sin\alpha\ .
\label{axisangle}
\eea
The orientation dependent part is then given by
\bea
M_{ij}P_{ij}(r,k)&=&\left(1+\cos\alpha-(1-\cos\alpha)\frac{(n\cdot r)^2}{r^2}\right)(rk)^2 \nonumber \\
&&\mathop{-}\,(1-\cos\alpha)\left(3\frac{(n\cdot r)^2}{r^2}-1\right)(rk+1)\ .
\eea

\begin{figure}[h!]
\centering
\includegraphics[scale=1.1]{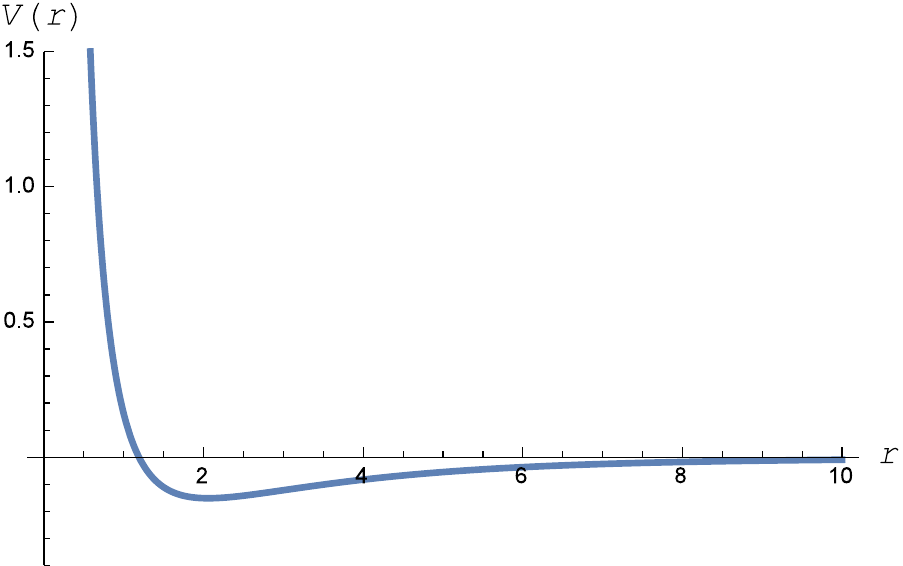}
\caption{Attractive channel potential, as a multiple of $N/\Lambda$. We can note the existence of a  minimum at  $x = 2.059$ and  $V_{\rm min}=-0.152$.}
\label{radial}
\end{figure}

We need a negative contribution from the dipole part to contrast the monopole part. Our best bet is to choose $r$, along with $\alpha$ and $n$, such that we get a positive contribution from $M_{ij}P_{ij}$, as that would mean that the long range force mediated by the pion is attracting the two objects, in contrast to the potential. We then choose the configuration of \textit{phase opposition}, where $r$ and $n$ are orthogonal and $\alpha$ indicates a half rotation: we can choose $r_i=(R,0,0)$, $n_i=(0,0,1)$ and $\alpha=\pi$, which corresponds to $M_{ij}P_{ij}=2rk+2$. This leads to $B^\dagger C=\pm i\sigma_3$. We will choose $B=\mathbf 1$ and $C=i\sigma_3$ as the phase opposition configuration (numerical analysis  confirms that the global minimum is attained in phase opposition). The potential in this channel is plotted against the distance between the two instantons $R$ in Figure \ref{radial}. We also calculate the asymptotic behaviors of the potential in the $r\to0$ and $r\to\infty$ limits, which are given by
\bea
&&V(r,i\sigma_3)\to \frac{4}{r^2}\frac{N_c}{\Lambda}\quad\text{ for }r\to0,\\
&&V(r,i\sigma_3)\to-\frac{48}{5r^3}\frac{N_c}\Lambda\quad\text{ for }r\to\infty.
\eea
The behavior for $r\to\infty$ is extracted by considering only the pion exchange interaction (which is the leading one when $r\to\infty$, as it is long range), while the behavior for $r\to0$ is considered by evaluating the monopole potential $\eqref{monoevaluate}$ and neglecting the gravitational warp: this is a standard problem of interaction for point charges in flat 4-dimensional space, with charges given by the first line of \eqref{motlinmon} with $H(z)=1$, while the dipole part of the interaction cancels.

We confront our potential with the potential obtained in \cite{Kim:2008iy}, through the consideration of an effective QFT of fermions (representing baryons) exchanging bosons (the mesons), which is obtained from the SS model.\footnote{Note that there is a normalization difference for the functions $\psi_n,\phi_n$: in the cited article $(\psi_n,\psi_m)=\delta_{nm}\mathop{=}<\phi_n,\phi_m>$. 
The correct identifications to make are then (LHS normalized as in this chapter, RHS normalized as in the cited article) $
\frac{1}{c_n}=\psi_n(0)^2$, $\frac1{d_n}=\psi'_n(0)^2$.}
We see that the two potentials look identical, apart from a numerical coefficient of three in front of the dipole part: our dipoles are three times as strong as in \cite{Kim:2008iy}. The reason for this difference will be clarified in Section \ref{confronto}.

\subsection{Binding energies and classical nuclei}

In the attractive channel, the potential (sketched in Figure \ref{radial}) assumes a minimum at $R_0=2.059$, of value $V_{\rm min}=-0.152 N_c/\Lambda$. The classical energy in the $\B=2$ sector is then given by
\bea
E_{2,c}=2M-0.152 \frac {N_c} \Lambda.
\eea
The classical energy is of order $N_c$, as expected. If $\Lambda\to\infty$, we get weakly bound  baryons of small size ($ {\cal O} (\lambda^{-1/2} )$) and large distance ($ {\cal O} (\lambda^{0} )$), large with respect to their size. This is exactly the limit where our computation is reliable.

We confront the value of $E_{2,c}$ with the classical energy of the $\B=1$ sector, $E_{1,c}=M$, by calculating the classical binding ratio that is independent of $N_c$. We have
\bea
{\rm BR}_{2,c}=\frac{2E_{1,c} - E_{2,c}}{2E_{1,c}}= \frac{0.608}{2.921\Lambda+\Lambda^2}.
\eea
As this quantity is always positive, for every value of $\Lambda$ and for every value of $N_c$, the classical deuteron turns out to be bound.

The experimental value of the binding ratio is 
\bea
{\rm BR}_{2, {\rm  exp}}=\frac{m_{p}+m_n - m_d}{m_{p}+m_n}=1.2\cdot10^{-3},
\eea
where $m_d = 1875.6$ MeV is the deuteron mass and $m_p$ and $m_n$ are the proton and neutron masses.
For a first, crude comparison with the SS model, we choose $\Lambda$ as in \cite{Sakai:2004cn} to fit the experimental value of the pion decay constant, and $\rho$ mass: $\Lambda=1.569$ (this corresponds to $\lambda = 16.632 $). 
With this value of $\Lambda$ we get ${\rm BR}_{2,c}=-0.086$, two orders of magnitude greater than the experimental value. Overestimating the binding energy is quite common also in the other Skyrme models. For the moment we make two preliminary comments on that. First, the extrapolation of our calculations to the phenomenological parameters contains many errors, mostly from $1/{N_c}$ and $1/\Lambda$ corrections which are not small. 
Second, the holographic model can be tuned to reach the correct order of magnitude for the binding energy by increasing $\Lambda$, at the price of loosing the fit with mesonic observables. The $\Lambda \to \infty$ limit, where the previous computation is valid, is in fact a weakly bound model.

\begin{figure}[h!]
\centering
\includegraphics[width=\linewidth]{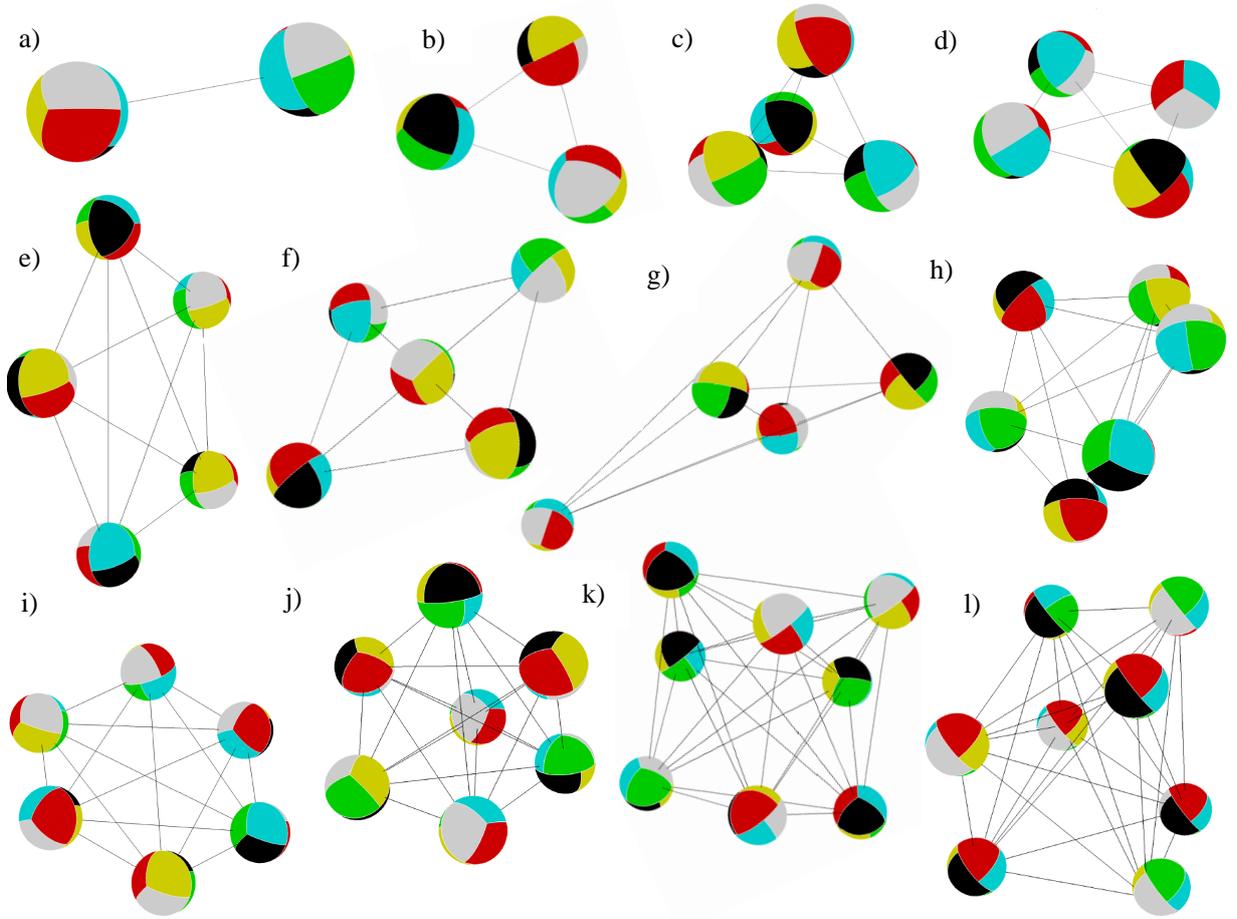}
\caption{Geometric configurations for stable and meta-stable nuclei  up to $\B = 8$. 'a-l)' denote the following respectively: deuteron, triangle, tetrahedron, square, pentagon, cross, tetrahedron with a satellite, two triangles, hexagon, tetrahedron with a triangle, two twisted squares and two rectangles. Colors represent the isosurface orientations, $ \upvarphi=(\varphi_{1},\varphi_{2},\varphi_{3})$, as radially projected onto the sphere which stands for the single charge instanton. The coloring scheme is as follows: Red/Green $= (\pm1,0,0)$, Cyan/Yellow $=(0,\pm1,0)$, White/Black $=(0,0,\pm1)$.}
\label{minima}
\end{figure}

\begin{figure}[h!]
\centering
\includegraphics[scale=.7]{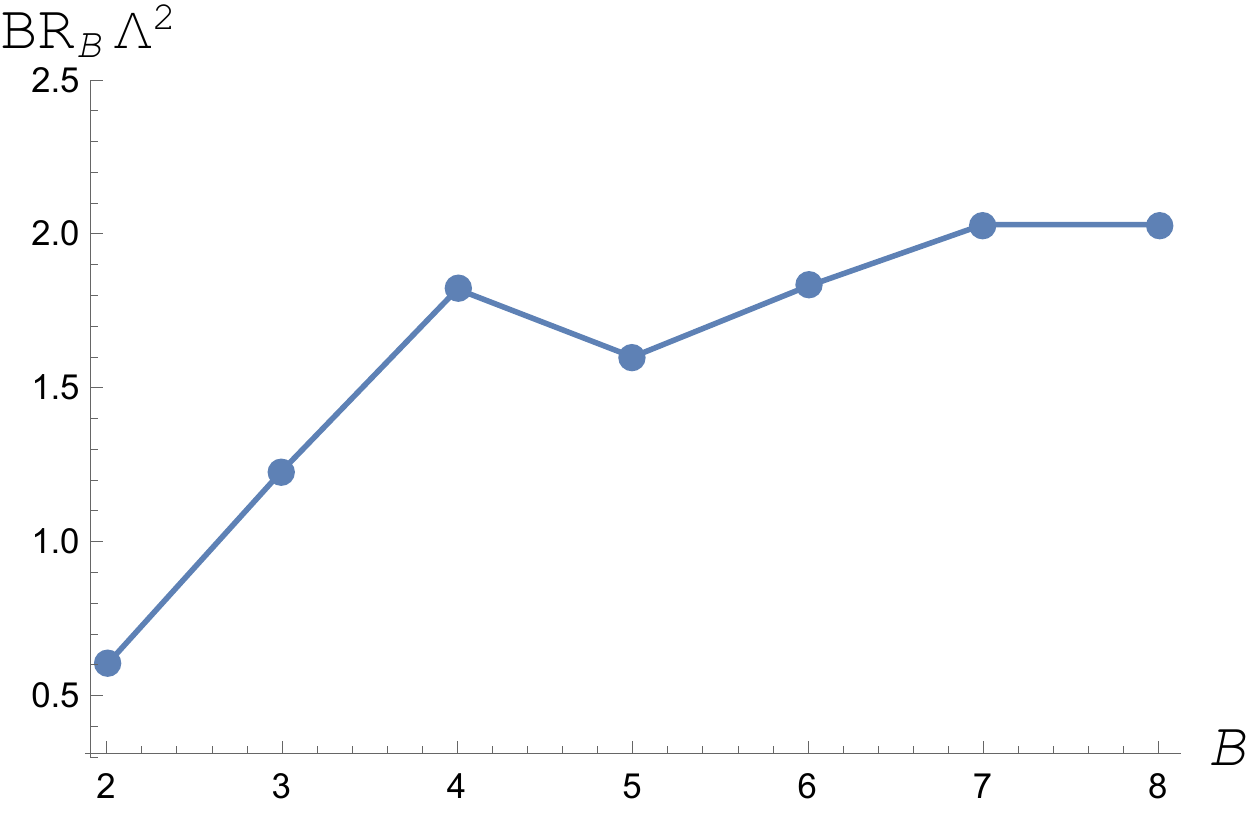}
\caption{Binding ratios (\ref{brat}) for the most stable nuclei up to $\B = 8$ for large $\Lambda$.}
\label{bindingratio1}
\end{figure}

We now use the potential to give some predictions about equilibrium configurations for nuclei with higher baryonic charge $\B$. Provided that the instantons are far away from each other, each of their core is localized in the linear zone of all the others. For $\B$ number of instantons, we define the potential $V_B$ as the sum of single potentials \eqref{potential} between all pairs of instantons, after which we find the minimum energy configuration.
We report the results of our analysis in Table \ref{manybody}, where we list the binding energies in different sectors and the different configurations numerically found for a stable solution.
\begin{table}[t!]
\centering
\begin{tabular}{c|c|c|c|c}
$\B$ & Shape & Details & $V_{\rm min}$& ${\rm BR}$\\
\hline
$2$ & Line & Distance = $2.06$ & $-0.15190$ & $0.61$\\
\hline
$3$ & Equilateral triangle & Side = $2.06$ & $-0.45570$ & $1.23$ \\
\hline
$4$ & Tetrahedron & Side = $2.06$ & $-0.91141$ & $1.82$\\
& Square & Side = $1.90$ & $-0.86988$ & $1.74$\\
\hline
$5$ & Pentagon & Side(outer) = $2.12$  & $-1.00268$ & $1.6$ \\
& (Tetrahedron + & Tetrahedron (Base = $2.0$ , Sides = $2.3$),  & $-0.99127$ & $1.58$\\
& Satellite) & Satellite (increasing) = ($3.2,3.7,5.0$) & &\\
& Cross & Distance from the center = $2.16$ & $-0.94359$ & $1.50$ \\
\hline
$6$ & Square + 2 satellites & Tetrahedron = ((base: $2.15$),$2.02,3.26$), & $-1.3746$ &$1.83$\\
& & Distance from the two satellites = & & \\
& & ((to base: $3.52,3.74,3.34,3.82$), & &\\
& & (to roof: $2.06,3.3.15$)) & &\\
& Hexagon & Side (outer) = $2.02$ & $-1.3627$ & $1.81$\\
\hline
$7$ & Tetrahedron & Pyramid Base = ($1.98$,$2.35$,$2.67$,$3.26$), & $-1.7740$ &$2.03$\\
& + triangle & Distance to the satellites = & & \\
& & ($2.34,2.34,3.55,3.55$) & & \\
& & ($3.25,3.25,3.55,3.55$) & & \\
& & ($1.98,1.98,3.70,3.70$) & & \\
\hline
$8$ & Two twisted squares & Square sides = (2.34,3.85)& $-2.0326$ & $2.03$\\
& & Twisted node connections(increasing) =&\\
& & (2.63,3.19,3.72,4.51) &\\
& Two rectangles & Rectangles:  & $-1.9632$ & $1.96$\\
& &Base = (2.03,4.45), Roof =(2.38,4.04)&\\
& & Distance between closest nodes =&\\
& &(3.49,3.57,3.49,3.57)&
\end{tabular}
\caption{Multi-instanton configurations for stable and meta-stable nuclei up to $\B=8$, details of the configuration shape, along with the potential and the binding ratios which are given in units excluding the lambda dependence.}
\label{manybody} 
\end{table}
In Figure \ref{minima} we diagrammatically show the multi-instanton configuration results for the stable and meta-stable nuclei, up to $\B=8$. For $\B=3$ there is a unique solution, the equilateral triangle.  For $\B \geq 4$, we find multiple local minima. In Figure \ref{bindingratio1} we plot the classical binding ratios as a function of $\B$ for the preferred configurations: 
\bea
\label{brat}
{\rm BR}_{B}=\frac{ \B E_{1,c} - E_{\B,c}}{B  E_{1,c}} \ .
\eea

\section{Zero-mode quantization and the Deuteron}
\label{quattro}

We begin by reviewing the effective zero-mode Lagrangian and its quantization  for the Sakai-Sugimoto instanton in the $\B=1$ sector. The moduli space is $\mathcal M=\mathbb R^3\times SU(2)$
with the metric (\ref{metsing}). 
The zero-mode lagrangian is then given by (in unscaled units)
\bea
L=\frac{1}{2}M\dot X^i\dot X^i+\frac14M\mu^2\omega_{L,i}\omega_{L,i}-M\ ,
\eea
where $\omega_{L,i}$ represent the left invariant (body fixed) angular velocities on $SU(2)$, ${\omega_L}_i=-i\tr[G^\dagger\dot G\sigma_i]$. We could have used the right invariant (space fixed) angular velocities ${\omega_R}_i$, due to the fact that ${\omega_L}_i{\omega_L}_i={\omega_R}_i{\omega_R}_i$. We discuss, in Appendix \ref{appa}, the role of the left and the right invariant velocities into detail. We have the same Lagrangian of a rigid body. We define canonical momenta
\bea
P^i=M \dot X^i\ ,\quad\quad J_i=\frac12M\mu^2\omega_{L,i}\ ,
\eea
and write the Hamiltonian as
\bea
H=\frac{P^iP^i}{2M}+\frac{J_iJ_i}{M\mu^2}+M\ .
\eea
As $J_i$ is the body fixed angular momentum, we can define the space fixed angular momentum by
\bea
I_i=-M(G)_{ij}J_j.
\eea
Among angular momenta, we have the commutation rules
\bea
 [I_i,I_j]=i\epsilon_{ijk}I_k,\quad [J_i,J_j]=i\epsilon_{ijk}J_k,\quad[I_i,J_j]=0.
\eea
We impose canonical commutation relations
\bea
[X^i,P^j]=i\delta^{ij}\ , \quad\quad[G,J_i]=iG\frac{i\sigma_i}{2}\ ,\quad\quad [G,I_i]=-i\frac{i\sigma_i}{2}G,
\eea
with all other commutators vanishing, we then write the generic ket state as
\bea
|\psi\rangle = |p^i,j,m_L,m_R\rangle,
\eea
with $p^i$ the momentum, $m_L$ is the eigenvalue of $J_3$ (to be interpreted as the spin), $m_R$ of $I_3$ (to be interpreted as isospin) and $j$ of $J^2$. In the rest frame, $p^i=0$, the energy eigenvalues are given by
\bea
E_{1}=\frac{j(j+1)}{M\mu^2}+M=\frac{N_c\Lambda}{8}+\sqrt{\frac{2}{15}}N_c+\sqrt{\frac56}\frac{j(j+1)}{N_c}\left(\frac{1}{8}-\sqrt{\frac65}\frac{64}{3\Lambda}\right)+o\left(\frac1{\Lambda^2}\right) .
\label{massformula}
\eea
As $M\propto N_c$,  quantum corrections due to the spinning are subleading, of order $N_c^{-1}$, and become negligible when $N_c\to\infty$, keeping $j$ fixed.
The proton is identified as the particle with isospin up, while the neutron has isospin down with $j=1/2$. States with an higher value of $j$ (always being semi-integer) give heavier baryons: as an example, we identify states with $j=3/2$ with the $\Delta$ states. States with $j$ integer are to be excluded by FR constraints, as we discuss in Appendix \ref{appb}.

We want to do the same for the $\B=2$ sector. For this we first need to study the zero-mode manifold, find its topology and metric, and then quantize it.

\subsection{The zero-mode manifold for $\B=2$}

We want to identify the manifold of the zero modes (which we call $\mathcal Z$), defined as a subspace of the twelve dimensional space we have, $\mathcal M_2$, (parametrized by the coordinates $(X_1,B,X_2,C)$) and on which the potential assumes a constant value.
We indicate an instanton field, centered in $X$ and with standard iso-orientation by $A_I(x-X)$. In this notation, an arbitrary field of topological charge 2 can be expressed within the linear approximation as
\bea
BA_I\left(x-X_1\right)B^{\dagger}+CA_I\left(x-X_2\right)C^{\dagger}.
\label{m2}
\eea
The space $\mathcal M_2$ is defined as the set of field configurations of this form.

The symmetry group of the action is
\bea
\mathcal L=\mathbb R^3\times SU(2)_I\times SU(2)_J\times\mathcal P \ ,
\eea
where $\mathbb R^3$ is the group of spatial translations, $SU(2)_I$ is the global part of the gauge group, $SU(2)_J$ is the double covering of the rotation group $SO(3)$ and $\mathcal P$ is the parity operation that sends $x\to-x$ while keeping the holographic $z$ coordinate invariant. From now on, we will neglect the center of mass position, by removing $\mathbb R^3$ from the symmetry group.

 Let $\tA$ be any static gauge field. The continuous part of $\mathcal L$ acts on $\tA$ according to
\bea
\tA(x,z)\to U[M(E)^*\tA(M(E)^{-1}x,z)]U^\dagger\ ,
\eea
where $U\in SU(2)_I$, $E\in SU(2)_J$, $M$ is the usual transformation from $SU(2)$ to $SO(3)$ and $M^*$ is the pullback on the vector field (rotating the fields $A_i$ and leaving the field $A_z$ invariant). The parity operation acts on the fields as
\bea
\tA_i(x,z)\to \tA_i(-x,z)\ ,\quad\quad\tA_z(x,z)\to-\tA_z(-x,z) \ .
\eea
We want to explicitly apply the transformation to the configuration $A^p+A^q$. As the transformation properties of the core solution and the linear approximation are the same, we can just use the linear approximation fields. All calculations remain the same for the core regions.

We start from a certain minimum energy configuration
\bea
\tA_I(x)=A_I\left(x-\frac{R}{2}\right)+\sigma_3 A_I\left(x+\frac R 2\right)\sigma_3,
\label{start}
\eea
where we define $R=(R_0,0,0)$ and $R_0$ as the position of the minimum of the potential in the attractive channel. From the linear approximation, we study the action of $\mathcal L$ on the field $A_I(x-X)$. An $SU(2)_I$ transformation acts in the usual way:
\bea
A_I(x-X)\to UA_I(x-X)U^\dagger\ ,
\eea
while an $SU(2)_J$ transformation acts as
\bea
&&A_i(x-X)\to M(E)_{ij}A_j\left(M(E)^{-1}x-X\right) \ ,\nonumber \\ 
&&A_z(x-X)\to A_z\left(M(E)^{-1}x-X\right)\ .
\eea
We can manipulate the $SU(2)_J$ transformations by 
\bea
A_{i}(x-X,z)\rightarrow -2\pi^{2}\mu^{2}M_{ij}(\epsilon_{jml}\sigma_{l}\partial_{m}^M+\sigma_{j}\partial_{z'})G(M^{-1}x,z,M^{-1}MX,z')|_{z'=0} \ ,
\eea
where  $M=M(E)$.
After this transformation, the derivative $\partial_m^M$ is now with respect to $M^{-1}x$. Note that we have multiplied $X$ by the identity. Using the fact that $G$ only depends on $|x-x'|$, we can remove $M^{-1}$. We must then transform the derivative according to
\bea
\partial^M_m=\frac{\partial}{\partial(M^{-1}x)^m}=\frac{\partial M_{ka}(M^{-1}x)^a}{\partial (M^{-1}x)^m}\frac{\partial}{\partial x^k}=M_{km}\frac{\partial}{\partial x^k}=M_{km}\partial_k \ .
\eea
Then we substitute in the expression for $A_i$, obtaining
\bea
A_{i}(x-R,z)\rightarrow-2\pi^{2}\mu^{2}(M_{ij}M_{km}\epsilon_{jml}\sigma_{l}\partial_{k}+M_{ij}\sigma_{j}\partial_{z'})G(x,z,MX,z')|_{z'=0}.
\eea
We can use the fact that $\epsilon$ is an invariant tensor, $\epsilon_{ijk}M_{ai}M_{bj}M_{ck}=\epsilon_{abc}$, by substituting $\epsilon_{jml}M_{ij}M_{km}\sigma_{l}\partial_{m}=\epsilon_{ijk}M_{kl}\sigma_{l}\partial_{j}$. Then, we use $M_{ij}\sigma_j=E^\dagger\sigma_i E$ to obtain
\bea
A_{i}(x-X,z)\rightarrow E^{\dagger}A_{i}(x-M(E)X,z)E.
\eea
The action on $A_{z}$ is the same:
\bea
A_{z}(x-X,z)\rightarrow -2\pi^{2}\mu^{2}\sigma_{i}\partial_{i}^ML(M^{-1}x,z,0,0).
\eea
Working as before, we get
\bea
A_{z}(x-X,z)\rightarrow E^{\dagger}A_{z}(x-M(E)X,z)E.
\eea
Regarding parity, it is trivial to verify that (remembering that $\epsilon$ takes a minus sign for the parity operation)
\bea
A_{i}(x-X,z)\rightarrow
A_{i}(x+X,z)\ ,\:\:\:\:\:\:A_{z}(x-X,z)\rightarrow A_{z}(x+X,z)\ .
\eea
The action of the continuous part of $G$ on the fields is then
\bea
\t{A}_{I}(x,z)   \rightarrow &&UE^{\dagger}A_I\left(x-M(E)\frac{R}{2},z\right)(UE^{\dagger})^{\dagger}\nonumber \\ &&\mathop+U\sigma_{3}E^{\dagger}A_I\left(x+M(E)\frac{R}{2},z\right)(U\sigma_{3}E^{\dagger})^\dagger.
\label{trans}
\eea
Eventually, parity can be used to change the sign of $\frac{R}{2}$. 

We say that a field configuration $\mathcal A_I$ belongs to the zero-mode manifold, if it can be written as
\bea
\mathcal A_I=&&UE^{\dagger}A_I\left(x-(-)^PM(E)\frac{R}{2},z\right)(UE^{\dagger})^{\dagger} \nonumber\\&&\mathop+Ui\sigma_{3}E^{\dagger}A_I\left(x+(-)^PM(E)\frac{R}{2},z\right)(Ui\sigma_{3}E^{\dagger})^{\dagger},
\label{zero}
\eea
for some matrices $U$ and $E$, belonging to $SU(2)$, and with $P$, the parity eigenvalue (defined modulo $2$): this eigenvalue assumes only values $P=0$ and $P=1$. The coordinates on this manifold are then $(U,E,P)$. We can act on those coordinates by a left action and a right action on the matrices or by using parity, sending $P$ into $P+1$ (modulo $2$).

To complete the definition of the zero-mode manifold, we have to discuss the isotropy group of the action on the coordinates. To do that, we use the notation of \cite{Leese:1994hb}, where a similar analysis in the Skyrme model is given by: $\mathcal O_{ai}$ represents a right translation of $\pi$ of the matrices $U$ and $E$ around the $a-$th isospatial axis and $i-$th spatial axis, while $\mathcal P_{ai}$ represents the same action on $U$ and $E$, with a change of sign. The values of the indices for $\mathcal O$ and $\mathcal P$ go from $0$ to $3$, where $0$ represents no transformations performed. As examples, $\mathcal O_{02}$ is the transformation $(U,E,P)\to(U,Ei\sigma_2,P)$ and $\mathcal P_{13}$ is $(U,E,P)\to(Ui\sigma_1,Ei\sigma_3,P+1)$. In addition to such transformations, we also have two $\mathbb Z_2$ transformations: $(U,E,P)\to(-U,E,P)$ and $(U,E,P)\to(U,-E,P)$ that obviously leave $\eqref{zero}$ invariant. In the following sections, we will take the symmetries $\mathbb Z_2$ as intended everywhere, and neglect overall signs in the $(U,E)$ matrices. In this notation, the transformations
\bea
\mathcal H=\{1,\mathcal O_{11},\mathcal O_{12},\mathcal O_{03},\mathcal P_{30},\mathcal P_{21},\mathcal P_{22},\mathcal P_{33}\}
\eea
form a group that leaves $\eqref{zero}$ invariant, as can be verified easily. There are no left translations of the matrices $U$ and $E$ that leave $\eqref{zero}$ invariant.

The zero-mode manifold is then defined by quotienting the manifold
\bea
SU(2)_I\times SU(2)_J\times \mathcal P
\eea
by the stabilizer $\mathcal H$. Actually, as in \cite{Leese:1994hb}, this manifold is isomorphic to
\bea
\mathcal Z=SU(2)_I\times SU(2)_J/\{1,\mathcal O_{11},\mathcal O_{12},\mathcal O_{03}\}.
\label{zeroman}
\eea
We now prove this assumption. A class in $SU(2)_I\times SU(2)_J\times P/\mathcal H$ can be expressed by choosing a set $(U,E,P)$ and acting on it with all transformations of $\mathcal H$. We indicate such an equivalence class by $\{(U,E,P)\}$. A class in $\mathcal Z$ is obtained by taking a set $(U,E)$ and then acting with the stabilizer. We denote such a class as $\{(U,E)\}$. We define the function on $\mathcal Z$, given by $f(\{(U,E)\})=\{(U,E,0)\}$, and state that this function is an isomorphism. This function is surjective, since any class of the form $\{(A,B,0)\}$ can be obtained by applying $f$ onto $\{(A,B)\}$, and similarly for $\{(A,B,1)\}$ by applying $f$ onto $\{(Ai\sigma_3,B)\}$, while noting that $\mathcal P_{30}(Ai\sigma_3,B,0)=(A,B,1)$. To prove injectivity, we define two pairs of matrices $(A,B)$ and $(C,D)$, such as $f(\{(A,B)\})=f(\{(C,D)\})=\{(A,B,0)\}$. Then $\{(C,D)\}$ must be obtained from $\{(A,B)\}$ by acting with a transformation  $\mathcal O_{ai}$ without parity, so that $\{(A,B,0)\}=\{(C,D,0)\}$. With these two properties, the function $f$ is an isomorphism and therefore we can adopt $\mathcal Z$ as the zero-mode manifold for the $\B=2$ sector.

We now build a Lagrangian on this manifold. For each instanton, we derive its kinetic energy through the metric \eqref{metsing}. In our usual coordinates $(X_1,B,X_2,C)$, defining left invariant angular velocities as $\omega_{B,i}=-i\tr(B^{\dagger}\dot B\sigma_i)$ and analogously for $\omega_{C,i}$, we take the result from the $\B=1$ sector in order to write the metric as\footnote{We are here neglecting eventual corrections in the metric that come from the overlapping of the two single instanton fields, as they bring contributions to the energy of order $1/N_c\Lambda$ and are negligible in both limits.}
\bea
g|_{\mathcal M}=dX_1^idX_1^i+2\mu^2d\Omega_{SU(2),B}+dX_2^idX_2^i+2\mu^2d\Omega_{SU(2),C}.
\eea
The kinetic energy on $\mathcal M_2$ is then
\bea
T=\frac{1}{2}M\left(\dot X_1^i\dot X_1^i+\dot X_2^i\dot X_2^i+\frac12\mu^2\omega_{B,i}\omega_{B,i}+\frac12\mu^2\omega_{C,i}\omega_{C,i}\right).
\eea
We modify the spatial coordinates as usual, defining a center of mass coordinate, $r_i$, and a global translation, $X_i$. From now on, we will neglect global translations by redefining $\mathcal M_2$ through the coordinates $(r,B,C)$, which specify a field configuration through
\bea
BA_I\left(x-\frac r 2\right)B^{\dagger}+CA_I\left(x+\frac r 2\right)C^{\dagger}.
\label{m2cen}
\eea
The kinetic energy becomes
\bea
T=\frac{1}{4}M\left(\dot r^i\dot r^i+\mu^2\omega_{B,i}\omega_{B,i}+\mu^2\omega_{C,i}\omega_{C,i}\right).
\eea
We must embed $\mathcal Z$ into $\mathcal M_2$, finding a law that allows us to find the coordinates on $\mathcal M_2$ through the coordinates of $\mathcal Z$. The embedding law is obtained by confronting \eqref{trans} with \eqref{m2cen}:
\bea
\begin{cases}
r_i=M(E)_{ij}R_j,\\
B=UE^\dagger,\\
C=Ui\sigma_3E^\dagger.
\end{cases}
\label{imm}
\eea
To transform the kinetic energy in the zero-mode manifold, we need to transform the velocities. We define the (left invariant) angular velocities $\omega_i$ relative to the matrix $E$ and $\Omega_i$ relative to the matrix $U$. First, we compute the derivative $\dot M_{ij}(E)$. Inverting the relation and defining $\omega_i$, we get
\bea
E^{\dagger}\dot{E}=\frac{i}{2}\omega_{i}\sigma_{i}.
\eea
This can be used to compute
\bea
\dot{M}_{ij}\sigma_{j} &=&\dot{E}^{\dagger}\sigma_{i}E+E^{\dagger}\sigma_{i}\dot{E}
= E^{\dagger}\sigma_{i}EE^{\dagger}\dot{E}-E^{\dagger}\dot{E}E^{\dagger}\sigma_{i}E \nonumber \\&=& \frac{i}{2}M_{ij}\omega_{k}[\sigma_{j},\sigma_{k}]=M_{ij}\epsilon_{kjl}\omega_{k}\sigma_{l}.
\eea
This implies
\bea
\dot{M}_{ij}=\epsilon_{klj}M_{il}\omega_{k}.
\eea
In the following, we denote $O_i$ as the rotation by $\pi$ around the $i$-th axis, while $M$ is the usual $SO(3)$ matrix associated to $E$.
\bea
\begin{cases}
B^{\dagger}\dot{B}=E(U^{\dagger}\dot{U}-E^{\dagger}\dot{E})E^{\dagger}\implies\omega_{B,i}=M_{ij}(\Omega_{j}-\omega_{j}),\\
C^{\dagger}\dot{C}=E(\sigma_{3}U^{\dagger}\dot{U}\sigma_{3}-E^{\dagger}\dot{E})E^{\dagger}\implies\omega_{C,i}=M_{ij}(O_{3,jk}\Omega_{k}-\omega_{j}),\\
\dot{r}_{i}=\dot{M}_{ij}R_{j}=\epsilon_{jlk}M_{ik}\omega_{l}R_{j}.
\end{cases}
\eea
We obtain
\bea
&&\omega_{B,i}\omega_{B,i}+\omega_{C,i}\omega_{C,i}=2\omega_{i}\omega_{i}+2\Omega_{i}\Omega_{i}-2(\delta_{kl}+O_{3,kl})\omega_{k}\Omega_{l},\\\nonumber
&&\dot{r}_{i}\dot{r}_{i}=\epsilon_{jlk}M_{ik}\omega_{l}R_{j}\epsilon_{abc}M_{ia}\omega_{c}R_{b}=\omega^{2}R^{2}-(\omega\cdot R)^{2}=(\omega^{2}_{2}+\omega^{2}_{3})R_{0}^{2}.
\eea
The matrix $\delta_{kl}+O_{3,kl}$ has only one nonvanishing element, which has the indices $k=l=3$ and equals $2$. The kinetic energy in the zero-mode manifold $\mathcal{M}$ then becomes
\bea
\left.T\right|_{\mathcal{Z}}=\frac{1}{2}M\left(\mu^{2}\omega_{1}^{2}+\left(\mu^{2}+\frac{R_{0}^{2}}{2}\right)\omega_{2}^{2}+\frac{R_{0}^{2}}{2}\omega_{3}^{2}+\mu^{2}(\Omega_{1}^{2}+\Omega_{2}^{2}+(\Omega_{3}-\omega_{3})^{2})\right).
\eea
In the zero-mode manifold, the potential energy attains its minimum value, that is $V_{\rm min}$ plus the self-energies $2M$. The Lagrangian is then given by
\bea
L|_{\mathcal Z}=T|_{\mathcal Z}-V_{\rm min}-2M.
\eea
\subsection{The quantum deuteron: quantizing the zero-mode manifold}

We quantize the zero-mode manifold $\mathcal Z$ by calculating the conjugate momenta from $L|_{\mathcal Z}$: here we denote as $L_{i}$ the momenta obtained by taking the derivative with respect to $\omega_{i}$, while $K_{i}$ are obtained by doing the same with respect to $\Omega_{i}$. We have
\bea
&&L_1=M\mu^2\omega_1,\quad L_2=M\left(\mu^2+\frac{R_0^2}2\right)\omega_2,\quad L_3=M\left(\frac{R_0^2}2+\mu^2\right)\omega_3-M\mu^2\Omega_3;\nonumber\\&&K_1=M\mu^2\Omega_1,\quad K_2=M\mu^2\Omega_2,\quad K_3=M\mu^2(\Omega_3-\omega_3).
\eea
The Hamiltonian is then
\bea
H|_\mathcal Z=\frac1{2M}\left(X_{ij}L_iL_j+Y_{ij}K_iK_j+2Z_{ij}L_iK_j\right)+V_{\rm min}+2M \ .
\eea
On this manifold, the potential is constant. The $X,Y,Z$ matrices are given by
\bea
&&X=
\left(\begin{array}{ccc}
\frac1{\mu^2}&0&0\\
0&\frac{2}{ 2\mu^2 +R_0^2}&0\\
0&0&\frac2{R^2_0}\\
\end{array}\right),\\ \quad 
&&Y=
\left(\begin{array}{ccc}
\frac1{\mu^2}&0&0\\
0&\frac{1}{\mu^2}&0\\
0&0&\frac2{R_0^2} + \frac1{\mu^2}\\
\end{array}\right),\\
&&Z=\left(\begin{array}{ccc}
0&0&0\\
0&0&0\\
0&0&\frac2{R_0^2}\\\end{array}\right).
\eea

Quantization proceeds as usual. We define the left invariant momenta as $L_i$ and $K_i$, the right invariant momenta as $J_i=-M(E)_{ij}L_j$, where $I_i=-M(U)_{ij}K_j$ (with $L^2=J^2$ and $K^2=I^2$), and write the ket state as
\bea
\ket{\psi}=\ket{k,k_3,i_3,l,l_3,j_3}.
\label{finket}
\eea
The definition of the quantum numbers is straightforward.

Not all kets \eqref{finket} are to be considered physical states, due to the fact that the zero-mode manifold is defined discretely, as in \eqref{zeroman}. We discuss the FR constraints and the details of the quantization process in Appendix \ref{appb}. Here, we cite the result: the only states that are compatible with the FR constraints are
\bea
&&\ket{D}=\ket{0,0,0,1,0,j_3},\quad\ket{I_{0}}=\ket{1,0,i_3,0,0,0},\\\nonumber&&\ket{I_{1}}=\frac{1}{\sqrt{2}}(\ket{1,1,i_3,0,0,0}+\ket{1,-1,i_3,0,0,0}).
\eea
We note that $\ket{D}$ has the right quantum numbers to be identified as the deuteron state (isospin singlet and spin triplet). By direct evaluation of $H|_{\mathcal Z}$ on the states that we have found (through the use of an explicit representation of the $L_i$ and $K_i$), we discover that they are eigenvectors of the Hamiltonian, with eigenvalues
\bea
\nonumber H|_{\mathcal Z}\ket{D} &=& \left(\frac{1}{2\mu^{2}M}\left(1+\frac{1}{1+\frac{R_{0}^{2}}{2\mu^{2}}}\right)+2M+V_{\rm min}\right)\ket{D}=E_D\ket{D},\\\nonumber 
H|_{\mathcal Z}\ket{I_{0}} &=& \left(\frac{1}{\mu^{2}M}+2M+V_{\rm min}\right)\ket{I_0}=E_{I_0}\ket{I_0},\\ 
H|_{\mathcal Z}\ket{I_{1}} &=& \left(\frac{1}{\mu^{2} M}\left(1+\frac{\mu^{2}}{R_{0}^{2}}\right)+2M+V_{\rm min}\right)\ket{I_{1}}=E_{I_1}\ket{I_1}.
\label{rote}
\eea
The deuteron state turns out to be the lowest energy state, with the lowest rotational energy contribution to the Hamiltonian. Due to the presence of the factor $1/M$, we have that the rotational energies are of order $N_c^{-1}$ (as expected, since they are subleading) and $\Lambda^0$, such that they are relevant in the $\Lambda\to\infty$ limit.

\section{Massive Modes}
\label{cinque}

What we have done so far is to obtain the leading order solution at large $N_c$ and large $\Lambda$, where all massive modes are frozen to their minimum value and only the zero-mode classical dynamics are relevant. All sorts of different $1/{N_c}$ and $1/\Lambda$ corrections are triggered by considering the massive modes and the quantum corrections. The most important ones, at least for large $\Lambda$, are the ones that would be the zero modes for the BPST instanton but are lifted when the solution is embedded in the SS model. There are various kinds of corrections that we need to study. We shall see that in order to reach the phenomenological values of the relevant parameters, $N_c=3$ and $\Lambda =1.569$, these corrections are very important.

\subsection{Baryon mass formula}

We begin by reviewing the effects of the massive modes in the $\B=1$ sector.
The standard YM instanton has eight moduli: four spatial coordinates $(X^i,Z)$ and four other that identify the size and $SU(2)$ orientation of the solution $(\mu,G)$. 
Due to the symmetry of the configuration under $G\to-G$, the moduli space of the single instanton is given by
\bea
\mathcal M=\mathbb R^4\times(\mathbb R^4/\mathbb Z^2).
\eea
We choose $(X^i,Z)$ as space and $y^I$ as the iso-space coordinates. The metric on $\mathcal M$ is
\bea
g=dX^i dX^i+dZ^2+2dy^I dy^I.
\eea

In the Sakai-Sugimoto model, $Z$ and $\mu$ cease to be exact moduli but have a potential 
\bea
U(\mu,Z)=M_0+M_0\left(\frac{\mu^2}6+\frac{64}{5\Lambda^2\mu^2}+\frac {Z^2}3\right)\ ,
\label{singlepot}
\eea
where $\mu = \sqrt{y_I y_I}$ and we take the $Z$ dependence from \cite{Hata:2007mb}.
The total Lagrangian is then
\bea
L=\frac12M_0(\dot X^i \dot X^i+\dot Z^2)+ M_0 \dot y^I\dot y^I-M_0-\frac 13{M_0}  Z^2- \frac16 M_0 \mu^2-\frac{64}{5\Lambda^2}\frac{M_0}{\mu^2} \ ,
\eea
where we define
\bea
M_0=\frac {N_c\Lambda}{8}\ .
\eea

It is convenient to cast the isospin part of the previous Lagrangian into radial coordinates. To this end, we define the $a_I$ coordinates through $\mu a_I=y_I$. This way, $a_I$ represent a point on $S^3/\mathbb Z_2$. In this scheme, the metric becomes
\bea
g=dX^idX^i+dZ^2+2(d\mu^2+\mu^2da^Ida^I)=dX^idX^i+dZ^2+2(d\mu^2+\mu^2g_{S^3}),
\label{totmetric}
\eea
where $g_{S^3}$ represents the standard metric on $S^3$.  
By restricting to the zero-mode manifold, $Z=0$ and $\mu = \mu_0$, we recover the previously discussed metric (\ref{metsing}).

The Hamiltonian operator can be written as $ H = H_{S} +  H_{I} $ where $H_{S}$  is the Hamiltonian relative to the $(X^i,Z)$ coordinates,
\bea
H_{S}=-\frac{1}{2M_0}(\partial_i\partial_i+\partial_Z^2)+\frac{1}{3}M_0  Z^2 \ ,
\eea
while  $H_{I}$ is the relative Hamiltonian for the isospace part,
\bea
H_{I}=-\frac{1}{4 M_0}\left(\frac{1}{\mu^3}\partial_\mu(\mu^3\partial_\mu)+\frac{1}{\mu^2}\left(\Delta_{S^3}-\frac{256}{5\Lambda^2}M_0^2\right)\right)+ \frac{1}{6} M_0 \mu^2 \ ,
\eea
where $\Delta_{S^3}$ is the Laplacian operator on the 3-sphere. 
Neglecting the total momentum, we have that a baryon state can be identified by the quantum numbers
\bea
\ket{n_z,n_\mu,j,m_L,m_R}.
\label{finalket}
\eea
The energy levels are (from \cite{Hata:2007mb})
\bea
E(n_z,n_\mu,j)&=&M_0+\frac{2n_z+1}{\sqrt{6}}+\frac{2n_\mu+1}{\sqrt6}+\sqrt{\frac{(2j+1)^2}{6}+\frac{2}{15}N_c^2} \nonumber \\
&\simeq& M_0+\sqrt{\frac{2}{15}} N_c+ \sqrt{\frac{5}{6}}\frac{(2j+1)^2}{4 N_c}+\frac{2n_z+1}{\sqrt{6}}+\frac{2n_\mu+1}{\sqrt6} .
\eea
The proton and the neutron are the lowest energy states of the $j=1/2$ representation, with $n_\mu=n_z=0$. States with higher $n_z$ or $n_\mu$ can be classified as resonances of the proton and the neutron. When evaluated with $n_\mu=n_z=0$, the energy levels are the same as in \eqref{massformula} (apart from a different zero of the energy), so we recover the previous results of the analysis of the zero-mode manifold.

\subsection{Sliding minimum}
\label{slidingminimum}

We can repeat the whole calculation for the classical potential (\ref{potential}) by inserting  the generic values of $\mu_1,\mu_2,Z_1,Z_2$.
For this, we have to modify (\ref{sol1}) in order to account for the additional coordinates. The final result is
\bea
&&V(r,B^\dagger C,\mu_i,Z_i) = \phantom{\frac12}\nonumber\\ &&\nonumber=\frac{4\pi N_c}{\Lambda}\sum_{n=1}^{\infty}\frac{\psi_n(Z_1)\psi_n(Z_2)}{c_n}\frac{e^{-k_n r}}{r}-\frac{\pi}{16}N_c\Lambda\mu_1^2\mu_2^2\frac{\phi_0(Z_1)\phi_0(Z_2)}{\pi}\frac{1}{r^3}P_{ij}(r,0)M_{ij}(B^{\dagger}C)\\
&&\mathop+\frac{\pi}{16}N_c\Lambda\mu_1^2\mu_2^2\left(\sum_{n=1}^{\infty}\left(\frac{\psi_n(Z_1)\psi_n(Z_2)}{c_n}-\frac{\phi_n(Z_1)\phi_n(Z_2)}{d_n}\right)\frac{e^{-k_n r}}{r^3}P_{ij}(r,k_n)\right)M_{ij}(B^{\dagger}C) \ . \nonumber\\
\label{totalpot}
\eea
The total potential is then
\bea 
V_{\rm tot}=U(\mu_1,Z_1)+U(\mu_2,Z_2)+V(r,B^\dagger C,\mu_i,Z_i),
\label{totalpotential}
\eea
where $U$ is defined in \eqref{singlepot} and $V$ in \eqref{totalpot}. We write the potential in the schematic form
\bea
 V_{\rm tot} &=& N_c\left(\frac{\Lambda}{48}(\mu_1^2+\mu_2^2)+\frac{8}{5 \Lambda}\left(\frac1{\mu_1^2} +\frac1{\mu_2^2}\right)+\frac{\Lambda}{24} (Z_1^2+Z_2^2) \right. \nonumber \\
&& \left.\mathop+\frac{1}{\Lambda}m(r) + \Lambda\mu_1^2\mu_2^2d(r,B^\dagger C)\right) \ , \label{tomin}
\eea
where $m(r)$ and $d(r,B^\dagger C)$ are the monopole and the dipole parts of the potential.

To look for the minimum we make the following ansatz: setting $Z_1=Z_2=0$, 
we restrict to the attractive channel: $r=(x,0,0)$ and $B^\dagger C=i\sigma_3$. We also restrict to the line $\mu = \mu_1=\mu_2$, while leaving $x,\mu$ as free variables to minimize.

As $m>0$ and $d<0$ when evaluated on the $x$ axis in phase opposition,
we have that as $\mu\to\infty$, $V_{\rm tot}\to-\infty$: this would
imply that the nucleons within the deuteron are stabilized when they
have infinite size. This is clearly an unphysical result that we can
exclude, since $\mu \to \infty$ is well outside the range of validity for the linear approximation. 
Depending on the values of $\Lambda$, a minimum can exist for finite $\mu$. We will look for that local minimum and neglect the clearly unphysical behavior of the potential for great values of $\mu$.

Taking the derivative of the potential $V_{\rm tot}$ within the ansatz, the minima for $x_0$ and ${\mu}_0$ are given by
\bea
&& m'(x_0)+\Lambda^2{\mu}_0^4  d'(x_0)=0 \ ,\\
&& \frac{\Lambda}{48} {\mu}_0-\frac{8}{5 \Lambda}\frac{1}{{\mu}_0^3}+\Lambda{\mu}_0^3d(x_0)=0 \ .
\eea
Substituting the first relation into the second we get
\bea
\frac{m'(x_0)}{d'(x_0)}\frac{1}{48}+\frac85=\frac{d(x_0)}{\Lambda}\left( -\frac{m'(x_0)}{d'(x_0)}\right) ^\frac32.
\label{mincond}
\eea
The minimum $x_0$ must then solve \eqref{mincond} in order to be a stationary point. Such an equation has to be solved numerically due to the non trivial $\Lambda$ dependence and the minimum point $x_0$ is not guaranteed to exist for any value of $\Lambda$. We report the graph showing the minimum as function of $\Lambda$ in Figure 8.
For $\Lambda\geq70$, the minimum always exists. 
In fact, setting $\Lambda=\Lambda_S = 1.569 $ in the above equation, numerical computation shows that the local minimum does not exist.
In the case $\Lambda\to\infty$, the equation reduces to
\bea
m'(x_0)=-\frac{384}{5}d'(x_0),
\eea
which can trivially be solved. We get $x_0\simeq2.06$, as in the previous case, and $\mu_0=0$, as expected from the previous case in the limit $\Lambda\to\infty$.

\subsection{Quantization in the harmonic approximation}
\label{massivethree}

We now extend our quantization of the $\B = 2$ sector to the massive modes. We start with the massive modes in $\mathcal M_2$, that correspond to having the instantons moving away from the phase opposition configuration. To do that, it is convenient to use coordinates $(r,B,C)$.

To perform this approximation, we calculate the second derivatives of the potential with respect to the coordinates. The derivatives with respect to the spatial coordinates $r$ are the standard derivatives, but we need a coordinate representation of the matrices $(B,C)$ in order to be able to identify the numerical results for the derivatives. We choose coordinates through the exponential map
\bea
B=\exp\left(i B_i\frac{\sigma_i}{2}\right),\quad\quad C=\exp\left(i C_i\frac{\sigma_i}{2}\right).
\eea
$B_i$ and $C_i$ are real and unconstrained numerical coordinates. They have a finite range, but as we are interested in the small changes of $B_i$ and $C_i$, we do not need to specify the range. In those coordinates, the left invariant velocities are given by
\bea
\omega_{B,i}=-i\tr(B^{\dagger}\dot B\sigma_i)=\dot B_i,\quad\quad\omega_{C,i}=-i\tr(C^{\dagger}\dot C\sigma_i)=\dot C_{i},
\eea
After canonical quantization of the matrix coordinates $B$ and $C$, we recover the quantum commutation relations with the left invariant angular momenta
\bea
 [B_i,J_{B,j}]=i\delta_{ij}\quad\quad[C_i,J_{C,j}]=i\delta_{ij},
\eea
with $J_{B,i}=\frac14 M_0\mu^2\omega_{B,i}$ and analogously for $J_{C,i}$. These coordinates can be used as canonical coordinates and we can perform the small oscillation approximation in the standard way. Returning back to the Lagrangian, we perform the derivatives and set the coordinates to their equilibrium values, $r=(R_{0},0,0)$, $B_{i}=(0,0,0)$ and $C_{i}=(0,0,\pi)$. Calling $\eta_a$ the displacement from the equilibrium coordinates (with $a=1,...,9$), the approximated Lagrangian can be written as
\bea
L|_{\mathcal M_2}=\frac{1}{2}M_{ab}\dot\eta^{a}\dot\eta^b-\frac{1}{2}V_{ab}\eta^a\eta^b-V_{\rm min}-2M,
\eea
where the mass matrix $M_{ab}$ is the diagonal matrix of eigenvalues
\bea
\frac{1}{2}\left[M,M,M,M\mu^2,M\mu^2,M\mu^2,M\mu^2,M\mu^2,M\mu^2\right],
\eea
and $V_{ab}$ has been computed numerically and shown in Table \ref{potapp}.
\begin{table}
\centering
$V_{ab}=\left(
\begin{array}{ccccccccc}
 0.142 & 0 & 0 & 0 & 0 & 0 & 0 & 0 & 0 \\
 0 & 0 & 0 & 0 & 0 & 0 & 0 & 0 & 0 \\
 0 & 0 & 0.662 & 0 & -0.681 & 0 & 0 &- 0.681 & 0 \\
 0 & 0 & 0 & 0 & 0 & 0 & 0 & 0 & 0 \\
 0 & 0 &- 0.681 & 0 & 0.701 & 0 & 0 & 0.701 & 0 \\
 0 & 0 & 0 & 0 & 0 & 0.542 & 0 & 0 & -0.542 \\
 0 & 0 & 0 & 0 & 0 & 0 & 0 & 0 & 0 \\
 0 & 0 &- 0.681 & 0 & 0.701 & 0 & 0 & 0.701 & 0 \\
 0 & 0 & 0 & 0 & 0 & -0.542 & 0 & 0 & 0.542 \\
\end{array}
\right)\dfrac{N_c}{\Lambda}$
\caption{Potential matrix at the equilibrium position.}
\label{potapp}
\end{table}
Solving the secular equation $\det(\omega^2 M_{ab}-V_{ab})=0$ we obtain three nonzero frequencies, as expected:
\bea
&&\omega_1=\frac {1.509} \Lambda, \quad\quad\omega_2=\frac{1.407}{\sqrt\Lambda} +o\left(\frac{1}{\sqrt{\Lambda^3}}\right),\\\nonumber
&&\omega_3=\frac{1.600}{\sqrt\Lambda}+o\left(\frac{1}{\sqrt{\Lambda^3}}\right).
\label{frequencies}
\eea
We can identify $\omega_1$ with the radial oscillation, which allows the constituents of the deuteron to vibrate along the axis joining them. Such an interpretation is suggested by its $\Lambda$ dependence, as the translational mode of inertia is proportional to $\Lambda$ and all entries in the $V$ matrix are multiplied by $\Lambda^{-1}$, giving an overall $\Lambda^{-2}$ dependence to the squared frequency. The other two frequencies are relatively small and non-global iso-rotations of the two objects, which do cost energy. The $\Lambda^{-\frac{1}{2}}$ dependence of the leading order comes from the fact that the moment of inertia has a leading order that is proportional to $M\mu^2$, which in turn is proportional to $\Lambda^0$, providing an overall $\Lambda^{-1}$ dependence to the squared frequencies.

The quantum Hamiltonian is readily written. We also include the contribution from the zero modes.
\bea
H|_{\mathcal M_2}=\sum_{i=1,2,3}\omega_i\left(a^\dagger_ia_i+\frac 1 2\right)+H|_{\mathcal Z}+V_{\rm min}+2M.
\eea
The ground states of the oscillators then give a contribution to the energy of the deuteron, energy of which is given by
\bea
E_{0,{\rm approx}}=\frac{\omega_1+\omega_2+\omega_3}{2}+V_{\rm min}+2M+\frac{1}{2\mu^{2}M}\left(1+\frac{1}{1+\frac{R_{0}^{2}}{2\mu^{2}}}\right).
\eea
All these terms have different $N_c$ and $\Lambda$ dependence. In particular we see that the two limits $N_c \to \infty$ and $\Lambda \to \infty$ do not commute. 
In order for our approximation to be valid we need to impose that the massive energies $\omega_{1,2}$ do not exceed the classical binding energy $V_{\rm min}$ and this is $N_c \gg \sqrt{\Lambda}$. In this way the two baryons are locked in the attractive orientation channel and rotation occurs only in the zero modes sub-manifold.

\subsection{Holographic massive modes}
\label{massivequattro}

In this section, we extend our harmonic approximation to the remaining two degrees of freedom, which in our approximation are also approximate moduli for the $\B=1$ instanton: the size, $\mu$, and the holographic coordinate, $z$. In the $\B=2$ sector, we have four additional massive modes for the deuteron, among which, the pairs that correspond to the same coordinate for different instantons are equivalent due to symmetry. 
In this case, we are using the total potential \eqref{totalpotential}, which includes the instanton self energies due to oscillations in their size and their position along $z$. The potential eigenvalues are estimated only approximately at the equilibrium position, keeping the radial distance for the minimum of the interaction potential and inserting the generic expression for the instanton size with both the $\Lambda$ and the $N_c$ dependences. For the remaining coordinates, the equilibrium positions are the same as before, in the attractive channel and at $z=0$.

We repeat the same set of calculations from the previous section by solving the determinant equation for the massive-mode frequencies and expanding them in order to observe the $\Lambda$ dependence. The mass matrix now includes four extra eigenvalues, two of which belong to the holographic coordinate are two times the position eigenvalues, for they are independent but not relative as opposed to radial coordinates, while the size eigenvalues are four times larger due to the factor of two, as can be read from \eqref{totmetric}. A total of seven nonzero frequencies are found as shown below

\bea
&&\omega_1=\frac {1.509}{\Lambda}\quad\quad\omega_2=\omega_3=0.816+o\left(\frac{1}{\Lambda^2}\right)\quad\quad\omega_4=\frac{2.813}{\sqrt{\Lambda}},\\\nonumber
&&\omega_5=\frac{3.200}{\sqrt{\Lambda}}+o\left(\frac{1}{\sqrt{\Lambda^3}}\right)\quad\quad\omega_6=\omega_7=0.816+o\left(\frac{1}{\Lambda}\right).
\label{frequencies2}
\eea 

Following the analysis of the previous section, we obtain the same frequency for oscillations in the radial distance along with the two angular frequencies, given by the $1/\sqrt{\Lambda}$ dependences, which have changed in magnitude due to the factor of $\mu^2$ in front of the angular metric. The remaining four frequencies, which differ from each other only in pairs and through the sign of their second-order term have a leading $\Lambda^0$ dependence due to the extra $\Lambda$ scaling in the total potential \eqref{tomin}. This particular scaling is due to the fact that these additional modes are coming solely from the single instanton energies at this order of approximation.   
  
Following the observation of the $\Lambda^0$ dependence from the additional massive mode frequencies, it can easily be verified that the energy contribution from these modes do not contribute to the deuteron binding energy, simply because their contribution in the $\B=2$ sector cancels their counterpart from the corrections to single instanton masses. Consequently, the $\Lambda$ dependence for the deuteron binding energy is unaffected by the addition of the holographic massive modes in our approximation.       
  
\subsection{The expectation value of the classical potential}
\label{confronto}

We are now in a position to understand the origin of the factor of 3, which differs between our potential (\ref{potential}) and the one in \cite{Kim:2008iy}. First, we average the potential of interaction over the quantum space that is generated by the coordinates $(r,B,C)$. This allows us to go beyond the zero-mode manifold when taking the average.

We recall our usual definitions for the angular momenta, as shown in Appendix \ref{appa}: If we write a generic $SU(2)$ matrix in quaternionic coordinates as $A=a_4\mathbf 1+ia_i\sigma_i$, we can represent the angular momentum operators as
\bea
J_i=\frac i2(a_i\partial_4-a_4\partial_i-\epsilon_{ijk}a_j\partial_k),\quad\quad I_{i}=\frac i2(a_4\partial_i-a_i\partial_4-\epsilon_{ijk}a_j\partial_k).
\label{momdef}
\eea
Using the $a_I$ coordinates, we can write a state of spin $j=\frac12$ as a polynomial in $a_i$ of degree one, quantum numbers of which we recall:
\bea
&\braket{a_I|\frac12,\frac12,\frac12}=\frac1\pi(a_1+ia_2),\quad\braket{a_I|\frac12,\frac12,-\frac12}=\frac i\pi(a_4+ia_3),\\\nonumber&\quad\braket{a_I|\frac12,-\frac12,\frac12}=\frac i\pi(a_4-ia_3),\quad\braket{a_I|\frac12,-\frac12,-\frac12}=\frac1\pi(ia_2-a_1).
\eea
A ket in the $(r,B,C)$ space can be specified by a radial coordinate and two angular momentum eigenvalues. We choose to neglect the quantization of the coordinate $r$, limiting our quantum space to only the angular variables $(B,C)$. Wavefunctions can be written as
\bea
\braket{b_I,c_I|j,m_L,m_R,j',m'_L,m'_R}=\braket{b_I|j,m_L,m_R}\braket{c_I|j',m'_L,m'_R}
\eea
and must be antisymmetric under the exchange of the $b_I$ and $c_I$ coordinates. As an example, the (unphysical) state which describes the first baryon as a spin up neutron and the second, as a spin up proton is given by
\bea
\braket{b_I,c_I|\frac12,-\frac12,\frac12\frac12}=\frac{i}{\pi^2}(b_4+ib_3)(c_1+ic_2).
\eea
We can relate the coordinates $b_I,c_I$ to those of the zero-mode manifold $e_I,u_I$ by using the immersion law
\bea
\begin{cases}
B=UE^\dagger,\\
C=Ui\sigma_3E^\dagger.
\end{cases}
\eea
This way, we can quickly search for a state that has the right quantum numbers to be interpreted as the deuteron state. As an example, the wavefunction
\bea
\braket{b_I,c_I|\psi}=\psi(b_I,c_I)=\frac{1}{\pi^2}(b_4c_3-b_3c_4+b_1c_2-b_2c_1),
\eea
(which is odd under the exchange $b_I\to c_I$), can be written in the zero-mode manifold, using the immersion law, as
\bea
\psi(u_I,e_I)=\frac1{\pi^2}(2(e_3^2+e_4^2)-1).
\eea
This wavefunction in the zero-mode manifold has the quantum numbers $\ket{k,k_3,i_3,l,l_3,j_3}=\ket{0,0,0,1,0,0}$, as can be verified by explicit calculation, representing the momenta in the form \eqref{momdef} and with opportune relabeling. Thus, we take the deuteron state in the $(B,C)$ space to be
\bea
\braket{b_I,c_I|\mathcal D}=\psi_D(b_I,c_I)=\frac{1}{\pi^2}(b_4c_3-b_3c_4+b_1c_2-b_2c_1).
\eea
To take the average of the potential over the deuteron state, we recall the rules of integration on $SU(2)$. We can parametrize $SU(2)$ as $S^3$ through the quaternionic representation and then use spherical coordinates
\bea
a_1=\cos\theta_1,\quad a_2=\cos\theta_2\sin\theta_1,\quad a_3=\cos\phi\sin\theta_2\sin\theta_1,\quad a_4=\sin\phi\sin\theta_2\sin\theta_1,
\eea
with coordinate ranges $\theta_i\in[0,\pi)$ and $\phi\in[0,2\pi)$. In those coordinates, the standard volume form is given by
\bea
V(\theta_1,\theta_2,\phi)=(\sin\theta_1)^2\sin\theta_2,
\eea
with total volume $2\pi^2$, that is the surface of the 3-sphere of unit radius. The definition of the scalar product is therefore the standard one
\bea
\braket{\psi|\xi}=\int_{S^3}(\psi(a_I))^*\xi(a_I)V(\theta_i,\phi)d\Omega,
\eea
($d\Omega=d\theta_1d\theta_2d\phi$) with each coordinate integrated over its range. When we have two sets of angular coordinates, we just have to follow the same procedure with both sets.

We are now prepared to compute the average of the potential. To do that, we write it in the form \eqref{potential}. As we fix the radial coordinate $r$ to be $r=(R_0,0,0)$, the average does not affect the monopole part, while the following integral must be calculated for the dipole part:
\bea
\braket{\psi_D|M_{ij}(B^\dagger C)|\psi_D}=\int_{S^3\times S^3}\psi_D(b_I,c_I)^2 M_{ij}(B^\dagger C)V(b_I)V(c_I)d\Omega_Bd\Omega_C.
\eea 
Computing this integral involves a very long sequence of trivial integrations of trigonometric functions. The calculation has been performed using Mathematica, obtaining the result
\bea
\braket{\psi_D|M_{ij}(B^\dagger C)|\psi_D}=\frac13\left(
\begin{array}{ccc}
	-1 & 0  & 0 \\
	0  & -1 & 0 \\
	0  & 0  & 1
\end{array}\right).
\eea
The factor $1/3$ is exactly what is needed to match the potential in \cite{Kim:2008iy}. As the approach taken in the article for calculating the potential involves quantization, this result is expected and we reproduce the article's result as the expectation value of the quantum operator which corresponds to our potential, calculated through classical field theory.
We believe that the two approaches are both correct, but in two different regimes. The fact $1/3$ should emerge when $N_c \ll \sqrt{\lambda}$ and the two baryons cannot be considered locked in the attractive orientation channel.

\section{Conclusion}
\label{sei}

We have extended  the solitonic picture for baryons in the SS model to higher charge nuclei. 
Working in the limit $\Lambda\to\infty$, we can place the instantons at large spatial distance with respect to their sizes and compute the static energy of the theory at the leading order in $\Lambda^{-1}$. We interpret the difference between this energy and the energy of two separated instantons as a classical potential for the nucleon-nucleon interaction.  We have shown that this potential depends on the relative distance between and the relative orientations of the single instantons, much in the same way as it happens in the Skyrme model.  We have identified a maximally attractive channel by fixing the relative orientations and have shown the existence of a classical bound state, computing the potential between the two objects and their binding energy. We have solved for the bound states of also the higher charge nuclei for up to $\B=8$.

The resulting picture of the two-instanton system is analogous to a rigid rotator, composed of two masses attached at a fixed distance, with a rotational degree of freedom that is interpreted as the classical spin and an internal degree of freedom that is interpreted as additional angular momentum, the isospin. 
Quantization forces this rotor to rotate. The ground state has exactly the quantum numbers to be interpreted as the deuteron state.

Our solution for holographic nuclear physics is valid at large $N_c$ and large $\lambda$.
In the large $N_c$ limit, the picture is entirely classical: more and more states arise from the quantization of the massive modes, until they form a continuum.
This picture is in agreement with large $N_c$ QCD.
In the large $\Lambda$ limit, the baryons shrink to zero size and the binding energy  goes to zero.
Extrapolating to physical values can be challenging. 
The linear approximation that we have used to calculate the potential is equivalent to keeping only the dominant terms in the $1/\Lambda$ expansion. As the physical value of $\Lambda$, as is extensively used in the literature (to fit the pion decay constant), is $\Lambda_{SS}=1.569$, we need, in principle, higher $1/\Lambda$ corrections to the interaction potential in order to be able to extrapolate numerical values, which can then be used to confront the physical/experimental data.

\appendix

\section{Spherical harmonics in four dimensions}
\label{appa}
In this section we recall the properties of the spherical harmonics in four dimensions. To do that, we first discuss the general problem of describing the motion of a particle in $\mathbb R^d$, equipped with the standard metric $\delta_{ij}$, with arbitrary $d$ and in the presence of a central potential $V(x^ix^i)=V(r)$. We then specialize to our case of interest; ($d=4$). In this section, we make no distinction between lowered and raised indices.

In standard Cartesian coordinates (and omitting all physical constants), the Lagrangian for the motion of such a particle is given by
\bea
L=\frac12(\delta_{IJ}\dot x^I\dot x^J)-V(r).
\eea
The corresponding quantum Hamiltonian is ($\Delta=\delta^{IJ}\partial_I\partial_J$)
\bea
H=-\frac12\Delta+V(r),
\eea
and the time independent Schr\"odinger equation for the wavefunction $F(X^I)$ is
\bea
-\frac12\Delta F(X^I)+(V(R)-E)F(X^I)=0.
\eea

We digress a little, by defining a harmonic function $H^{(L)}(X^I)$ as the solution of $\Delta H^{(L)}(X^I)=0$ and with the form of a complex homogeneous polynomial $X^I$ of rank $L$ (such as $H^{(L)}(t X^I)=t^LH(X^I)$). These functions can be written as
\bea
H^{(L)}(X^I)=C_{I_1I_2...I_L}X^{I_1}X^{I_2}...X^{I_L},
\eea
where summation is implied over the indices $I_j$, assuming values from $1$ to $d$ and $C$ is symmetric in all its indices. The condition $\Delta H^{(L)}=0$ puts constraints on the form of the complex coefficients $C_{I_1...I_L}$: This condition can be expressed as
\bea
C_{KKI_3...I_L}=0,
\label{sym}
\eea
that is, the trace of the first two indices of $C$ (or any couple of indices, due to symmetry) must vanish. To pave the way for spherical coordinates, we define $a^I$ through $X^I=ra^I$: $a^I$ will then be a unit vector in the $d$-dimensional space; an element of $S^{d-1}$.

We now switch to spherical coordinates, denoting the radius as $r$. Through the standard techniques of differential geometry, we have
\bea
\Delta=\frac{1}{r^d}\partial_r(r^d\partial_r)+\frac{\Delta_{S^{d-1}}}{r^2},
\eea
where $\Delta_{S^{d-1}}$ is the Laplacian operator on the unit sphere $S^{d-1}$. We define the spherical harmonics as
\bea
Y^{(L)}=r^{-L}H^{(L)}(X^I)=C_{I_1 I_2...I_L}a^{I_1}a^{I_2}...a^{I_L}.
\eea
Those functions are the eigenstates of $\Delta_{S^{d-1}}$, as
\bea
\Delta H^{(L)}=0=\frac{L(L+d-2)}{r^2}r^LY^{(L)}+\frac{1}{r^2}r^L\Delta_{S^{d-1}}Y^{(L)}.
\eea
We then obtain
\bea
\Delta_{S^{d-1}}Y^{(L)}=-L(L+d-2)Y^{(L)}.
\eea
We now count how many spherical harmonics of a certain rank can be built. This is done by counting the number of independent components from the tensor $C_{I_1...I_L}$, under the constraints of symmetry and tracelessness. The dimension of the space of symmetric tensors of rank $L$ over a $d$-dimensional space is
\bea
g_d(L)={{d+L-1}\choose{L}}.
\eea
We have to subtract a number of constraints, due to the requirement \eqref{sym}, which is a list of $g_d(L-2)$ equations. The final result is that, for any rank $L$, there are
\bea
{\rm deg}(L)=g_d(L)-g_d(L-2)
\eea
number of independent spherical harmonics. In the case of $d=3$, we get ${\rm deg}(L)=1+2L$, while for $d=4$, we get ${\rm deg}(L)=(1+L)^2$.

Returning to our original problem, we write the Laplacian in spherical coordinates and the wavefunction as $F(X)=R(r)Y^{(L)}$. This way, the angular dependence is completely solved and we obtain the equation for $R$:
\bea
-\frac12\frac{1}{r^d}\partial_r(r^d\partial_r R(r))+\frac{L(L+d-2)}{r^2}R(r)+(V(R)-E)R(r)=0.
\eea

We now set $d=4$ and study the spherical harmonics in four dimensions. In this case, we have
\bea
\Delta_{S^{d-1}}Y^{(L)}=-L(L+2)Y^{(L)},\quad\quad {\rm deg}(L)=(L+1)^2.
\eea
Henceforth, we will use the index convention $i=1,2,3$ and $I=1,2,3,4$. In order to study the representations, we consider a corresponding quantum problem: the motion of a test particle on $S^3$, described by the Lagrangian
\bea
L=\frac12 \dot a^I\dot a^I.
\label{basicla}
\eea
We define the canonical momenta as $\Pi_I=\dot{a}^I$. In the usual quantization scheme, we set $\Pi_I=-i\partial_I$. The quantum Hamiltonian is then
\bea
H=-\frac12\Delta_{S^3},
\label{lapham}
\eea
which is diagonalized by states such as $\braket{a^I|L,h}=Y^{(L)}(a^I)$, where $Y^{(L)}(a^I)$ is a spherical harmonic. $h$ is an index (or a set of indices) that takes $(L+1)^2$ different values, specifying the particular spherical harmonic that we intend to use.

In order to simplify our analysis and classify the irreducible representations, we use the isomorphism between $S^3$ and $SU(2)$, given by $U(a_I)=a_4\mathbf 1+i a_i\sigma_i$. On the $a^I$ coordinates, we can act with $SO(4)$, which leaves $S^3$ invariant.
The action of $SO(4)$ on $S^3$ can be expressed in terms of the left and the right transformations of $SU(2)$, acting on itself: taking $\ket{U}$ as a state that is centered on the $SU(2)$ matrix $U$, we can act through $L(P)\ket{U}=\ket{PU}$ and $R(P)\ket{U}=\ket{UP^\dagger}$, using any $SU(2)$ matrix $P$. For the trajectory $U(t)$ in $SU(2)$, we can define the left and the right invariant velocities as
\bea
\omega_{L,i}=-i\tr(U^\dagger\dot U\sigma_i),\quad\quad \omega_{R,i}=i\tr(\dot UU^\dagger\sigma_i)=-M_{ij}(U)\omega_{L,j},
\eea
where $M(U)$ is defined in \eqref{associated}. Through those velocities, we can connect the description in terms of $S^3$ and the description in terms of $SU(2)$. If $U=U(a^I)$, we have (using $a^Ia^I=1,\;a^I\dot a^I=0$)
\bea
\omega_{L,i}=2(\dot a_i a_4- a_i\dot a_4+\epsilon_{ijk}a_j\dot a_k),\quad\omega_{R,i}=2(a_i\dot a_4-\dot a_i a_4+\epsilon_{ijk}a_j\dot a_k),
\eea
and it is trivial to show that
\bea
\omega_{L,i}\omega_{L,i}=\omega_{R,i}\omega_{R,i}=4\dot a^I\dot a^I.
\label{conn}
\eea
We now substitute \eqref{conn} into the Lagrangian \eqref{basicla} and quantize the resulting Hamiltonian. We can use either the left invariant or the right invariant velocities in defining the momenta, getting respectively
\bea
J_i=\frac14\omega_{L,i},\quad\quad I_i=\frac14\omega_{R,i},\quad\quad H=2J_{L,i}J_{L,i}=2J_{R,i}J_{R,i}.\label{mome}
\eea
As in the literature, we call $J_i$ the body fixed angular momentum and interpret it as the spin, while calling $I_i$ the space fixed angular momentum and interpreting it as the isospin. The operators on the wavefunctions, which correspond to the momenta $I_i$ and $J_i$, must respect the following commutation relations, since the left and right invariant actions commute:
\bea
[J_i,J_j]=i\epsilon_{ijk}J_{k},\quad[I_i,I_j]=i\epsilon_{ijk}I_k,\quad[I_i,J_j]=0.
\eea
and, as $I_i=-M_{ij}(U)J_j$, we also have
\bea
I_iI_i=J_iJ_i=J^2.
\eea
(this is consistent with the definition of the Hamiltonian). We impose the fundamental commutator with the coordinate operator, $U$, by requiring that the left invariant momentum generates the right translations, while the right invariant momentum generates the left translations. 
\bea
[U,J_i]=iU\left(\frac{i\sigma_i}2\right),\quad\quad[U,I_i]=-i\left(\frac{i\sigma_i}2\right)U.
\eea

We classify the states through the eigenvalues of $J^2,J_3,I_3$, which are commuting operators. A ket state is then of the form $\ket{j,m_L,m_R}$, with $-j\leq m_L,m_R\leq j$ and $J^2\ket{j,m_L,m_R}=j(j+1)\ket{j,m_L,m_R}$: The states can be classified by the projections of two different angular momenta, sharing the total momentum eigenvalue $j$.

We can classify spherical harmonics by returning to the $a^I$ coordinates. In those coordinates, the quantization prescription is $\Pi_I=\dot a_I\to-i\partial_I$, such that the angular momenta become
\bea
J_i=\frac i2(a_i\partial_4-a_4\partial_i-\epsilon_{ijk}a_j\partial_k),\quad\quad I_i=\frac i2(a_4\partial_i-a_i\partial_4-\epsilon_{ijk}a_j\partial_k).
\eea
Comparing \eqref{lapham} and the Hamiltonian in \eqref{mome}, we have that $-4J^2=\Delta_{S^3}$. Acting with $J^2$ on a spherical harmonic, we get
\bea
J^2Y^{(L)}(a_I)=\frac L2\left(\frac L2+1\right)Y^{(L)}(a_I),
\eea
We see that the spherical harmonics of rank $L$ can be used as the wavefunctions of the $S^3$ representation for spin $j=L/2$.

We conclude with an example. For $L=1$, we get the representation $j=1/2$, where we have the four wavefunctions
\bea
&\braket{a_I|\frac12,\frac12,\frac12}=\frac1\pi(a_1+ia_2),\quad\braket{a_I|\frac12,\frac12,-\frac12}=\frac i\pi(a_4+ia_3),\\\nonumber&\quad\braket{a_I|\frac12,-\frac12,\frac12}=\frac i\pi(a_4-ia_3),\quad\braket{a_I|\frac12,-\frac12,-\frac12}=\frac1\pi(ia_2-a_1).
\eea
The reader can test these results, using the explicit form of the angular momenta in terms of the $a_I$, for the eigenvalues of the momentum operators.

\section{FR constraints and transformation properties}
\label{appb}
In this section we review the wavefunction representations and the transformation properties of the physical states we obtained. We have already given a wavefunction representation at the end of Appendix \ref{appa}, and here, we will provide a more formal representation that makes the transformation properties evident. We follow the conventions adopted in \cite{Leese:1994hb}.

A state in an $SU(2)$ representation is specified by the numbers
\bea
\ket\psi=\ket{j,m_L,m_R},
\eea
as in Appendix \ref{appa}. $SU(2)$ coordinates are expressed through the $SU(2)$ matrices, $U$: the coordinate wavefunction representation of $\ket\psi$ is given by
\bea
\braket{U|\psi}=\braket{U|j,m_L,m_R}=D^j_{m_L,m_R}(U),
\eea
where $D^j_{m_L,m_R}(U)$ is a Wigner matrix, and plays a role which is analogous to that of the spherical harmonics. The explicit form of the matrix $D$ depends on the coordinates that are used to describe the matrices $U$: We have seen at the end of Appendix \ref{appa} that the wavefunctions for $j=1/2$ are expressed in quaternionic coordinates. In this representation, the operators $J_i$ and $I_i$ act as expected:
\bea
&&\braket{U|J_3|j,m_L,m_R}=m_LD^j_{m_L,m_R}(U),\quad\braket{U|I_3|j,m_L,m_R}=m_RD^j_{m_L,m_R}(U),\quad\\\nonumber&&\braket{U|J^2|j,m_L,m_R}=j(j+1)D^j_{m_L,m_R}(U).
\eea
To act with rotations, we define the left translation operator, $L(P)$, as $L(P)\ket{U}=\ket{PU}$ and the right translation operator, $R(P)$, as $R(P)\ket{U}=\ket{UP^\dagger}$. On a $\ket{j,m_L,m_R}$ state, they act according to
\bea
&&\braket{U|L(P)|j,m_L,m_R}=D^{j}_{m_L,m_R}(P^\dagger U)=\sum_{m=-j}^jL(P)^j_{m_R,m}D^{j}_{m_L,m}(U),\label{left}\\
&&\braket{U|R(P)|j,m_L,m_R}=D^{j}_{m_L,m_R}(UP)=\sum_{m=-j}^jR(P)^j_{m,m_L}D^{j}_{m,m_R}(U).\label{right}
\eea
The explicit matrices $L(P)^j_{m_L,m_R}$ and $R(P)^j_{m_L,m_R}$ can be calculated in the following way: by letting $p_i$ be the exponential coordinates of $P$,
\bea
P=\exp\left(ip_i\frac{\sigma_i}{2}\right),
\eea
the matrices are given by
\bea
L^j_{m_L,m_R}(P)=\exp(ip_iI_i^{(j)})_{m_L,m_R},\quad\quad R^j_{m_L,m_R}(P)=\exp(ip_iJ_i^{(j)})_{m_L,m_R},
\eea
where the angular momentum matrices $I_i^{(j)}$ and $J_i^{(j)}$ form the spin $j$ irreducible representation of dimension $(2j+1)^2$.

In the $\B=1$ sector, we have a single matrix coordinate, the phase $G$, and a field configuration is written as
\bea
GA_I(x)G^\dagger.
\eea
A state is then simply given by
\bea
\ket{j,m_L,m_R}.
\eea
To quantize the instanton as a fermion we must have $\ket{G}\to-\ket{-G}$. As we have the relations $-G=\exp\left(i2\pi\frac{\sigma_3}2\right)G=G\exp\left(i2\pi\frac{\sigma_3}2\right)$ the constraint can be rewritten as
\bea
&&\exp\left(i2\pi J_3\right)\ket{j,m_L,m_R}=\exp\left(i 2\pi m_{L}\right)\ket{j,m_L,m_R}=-\ket{j,m_L,m_R};\\
&&\exp\left(i2\pi I_3\right)\ket{j,m_L,m_R}=\exp\left(i 2\pi m_{R}\right)\ket{j,m_L,m_R}=-\ket{j,m_L,m_R}.
\eea
The constraint then is implemented by selecting states with $m_L$ and $m_R$ half integer, such that only the states with $j$ half integer are physical. The states which correspond to $j=\frac12$ represent the basic nucleons, while the states which correspond to $j=\frac32$ represent the $\Delta$ states and so on. We can implement two types of transformations: iso-rotations, obtained by multiplying $G$ on the left ($G\to LG$) and transforming the wavefunction as in \eqref{left}, and rotations that are obtained by multiplying $G$ on the right ($G\to GR$) and transforming the wavefunction as shown in \eqref{right}. 

In the $\B=2$ sector, we define the zero-mode manifold as the manifold of the field configurations
\bea
BA_I(x-\frac{r}{2})B^\dagger+CA_I(x+\frac r2)C^\dagger,
\eea 
which can be written as
\bea
&&UE^{\dagger}A_I\left(x-M(E)\frac{R}{2},z\right)(UE^{\dagger})^{\dagger} \\\nonumber&&\mathop+Ui\sigma_{3}E^{\dagger}A_I\left(x+M(E)\frac{R}{2},z\right)(Ui\sigma_{3}E^{\dagger})^{\dagger}.
\eea
where the vector $R=(R_0,0,0)$ gives the bound state separation distance. We call $K_i$ and $L_i$ the left invariant momenta, relative respectively to $U$ and $E$ and $I_i=-M_{ij}(U)K_j$ and $J_i=-M_{ij}(E)L_j$ the respective right invariant angular momenta (obeying $K^2=I^2$ and $L^2=J^2$). We define those momenta to obey the rules
\bea
[K_i, U]=\frac i2 U(i\sigma_i),\quad\quad[L_i, E]=\frac i2 E(i\sigma_i),
\eea
from which
\bea
[I_i, U]=-\frac i2 (i\sigma_i)U,\quad\quad[J_i,E]=-\frac i2 (i\sigma_i)E,
\eea
follow. Among the momenta, we have commutation rules
\bea
[K_i,K_j]=i\epsilon_{ijk}K_k,\quad[L_i,L_j]=i\epsilon_{ijk}L_k,\quad[I_i,I_j]=i\epsilon_{ijk}I_k,\quad[J_i,J_j]=i\epsilon_{ijk}J_k,
\eea
such that a state in momentum space can be written as $\ket{\psi}=\ket{k,k_3,i_3,l,l_3,j_3}$.

The discrete symmetries we have are 
\bea
&&\mathcal O_{11}:(U,E)\to(Ui\sigma_1,Ei\sigma_1),\quad\mathcal O_{12}:(U,E)\to(Ui\sigma_1,E i\sigma_2),\\\nonumber&&\mathcal O_{03}:(U,E)\to(U,Ei\sigma_3),
\label{discr}
\eea
in addition to the symmetries $(U,E)\to (U,-E)$ and $(U,E)\to (-U,E)$. To impose FR constraints, we must assign to each closed path connecting those configurations, a phase of $\pm1$. For the last set of symmetries, this decision is an easy one. In the $\B=2$ sector, and in all sectors with $\B$ even, the wavefunction must be even under rotations of $2\pi$, either of the total spin or the isospin: $\ket{-U,E}=\ket{U,E}$ and $\ket{U,-E}=\ket{U,E}$. In opposition to the $\B=1$ sector, this implies that all momenta must have integer eigenvalues. We now take the symmetries \eqref{discr} into account.

We start by examining the symmetry $\mathcal O_{11}$. A path that starts at an arbitrary point $(U,E)$ and ends at $(Ui\sigma_1,E i\sigma_1)$ can be written as
\bea
U(\theta)=U\exp\left(i\theta \frac{\sigma_1}2\right),\quad E(\theta)=E\exp\left(i\theta \frac{\sigma_1}2\right),
\label{o11}
\eea
with $\theta\in[0,\pi]$. In terms of the coordinates $(B,C)$, this specific path corresponds to
\bea
B(\theta)=B,\quad C(\theta)=U\exp\left(i\theta \frac{\sigma_1}2\right)i\sigma_3\exp\left(-i\theta \frac{\sigma_1}2\right)E^\dagger.
\eea
We can see that, for $\theta=\pi$, we have $B(\pi)=B$ and $C(\pi)=-C$. This path corresponds to the $\mathcal Z_2$ symmetry which rotates a single instanton by $2\pi$. As we require the nucleon states to be fermionic states, the path $\mathcal O_{11}$ must be implemented as a noncontractible path: In coordinate space, we have $\ket{U,E}=-\ket{Ui\sigma_1,E i\sigma_1}$. We now examine the symmetry $\mathcal O_{12}$, which can be studied through the path
\bea
U(\theta)=U\exp\left(i\theta\frac{\sigma_1}{2}\right),\quad E(\theta)=E\exp\left(i\theta\frac{\sigma_2}2\right).
\label{o22}
\eea
This path is homotopic to the path \eqref{o11}, through the deformation ($s\in[0,\pi/2]$)
\bea
U(\theta,s)=U\exp\left(i\theta \frac{\sigma_1}2\right),\quad E(\theta,s)=E\exp\left(i\theta\left(\frac{\sigma_1}2\cos s+\frac{\sigma_2}2\sin s\right)\right).
\eea
When $s=0$, we have path \eqref{o11}, while when $s=\pi/2$, we have path \eqref{o22}. This means that we must impose $\ket{U,E}=-\ket{Ui\sigma_1,Ei\sigma_2}$. Lastly, $\mathcal O_{03}$ is trivially obtained by composing $O_{11}$ and $O_{12}$: in coordinate space, this means that $\ket{U,E}=\ket{U,Ei\sigma_3}$.

As the constraints are all expressed through the right multiplications, they  require determinate behavior of the physical states under the transformations generated by $K_i$ and $L_i$. The constraint $\ket{U,E}=-\ket{Ui\sigma_1,Ei\sigma_1}$ is implemented by writing
\bea
\ket{U(-i\sigma_1),E(-i\sigma_1)}=\exp\left(i\pi(L_1+K_1)\right)\ket{U,E}=-\ket{U,E}.
\eea
We see that a physical state $\psi$ must be an eigenstate of $\exp(i\pi(J_1+K_1))$, with the eigenvalue $-1$. Working in a similar way with the other constraints, we have
\bea
\exp\left(i\pi(L_1+K_1)\right)\ket\psi=\exp\left(i\pi(L_2+K_1)\right)\ket{\psi}=-\ket{\psi},\;\exp\left(i\pi L_3\right)\ket\psi=\ket\psi.
\label{transfr}
\eea
We see that the first two constraints exclude the state $l=k=0$, which is transformed trivially by all rotations. We search for physical states in the $(k=1,l=0)$ and the $(k=0,l=1)$ representations (which have the lowest angular momentum and hence are good candidates for the ground state): We do this explicitly by constructing the appropriate matrix representations and computing the associated matrices for the transformations, \eqref{transfr}. We obtain the compatible states
\bea
&&\ket{D}=\ket{0,0,0,1,0,j_3},\quad\ket{I_{0}}=\ket{1,0,i_3,0,0,0},\\\nonumber&&\ket{I_{1}}=\frac{1}{\sqrt{2}}(\ket{1,1,i_3,0,0,0}+\ket{1,-1,i_3,0,0,0}),
\eea
which can be written in wavefunction form as
\bea
\braket{U,E|D}=D^1_{0,j_3}(E),\;\braket{U,E|I_0}=D^1_{0,i_3}(U),\;\braket{U,E|I_1}=D^1_{1,i_3}(U)+D^1_{-1,i_3}(U).
\eea
We see that each ket among $\ket{D},\ket{I_0},\ket{I_1}$ represents a triplet of states, degenerate in energy and differing only by the projection of a right-invariant momentum. Those states transform into each other under full rotations and iso-rotations. Due to the definition of the $U$ and $E$ matrices in terms of single phases $B$ and $C$, we have that simultaneous left translations $(B,C)\to(LB,LC)$, which are interpreted as iso-rotations in the $\B=1$ sector, can be realized by the transformation $(U,E)\to(LU,E)$, while simultaneous right translations $(B,C)\to(BR^\dagger,CR^\dagger)$ represent total rotations, as can be realized by $(U,E)\to(U,RE)$. In either case, we transform the matrices $(U,E)$ by acting on them from the left. We then see that the states transform as in \eqref{left} and the rules for performing physical transformations on the zero-mode manifold are coherent with the FR constraints.

\section*{Acknowledgments}

We thank F.~Bigazzi, A.~Cotrone, D.~Harland, L.~Marcucci, P.~Sutcliffe, M.~Speight, M.~Viviani, A.~Wereszczynski, A.~Manenti for useful discussions.
S.~Bolognesi is founded by the program ``Rientro dei Cervelli Rita Levi Montalcini''.
S.~B.~G.~thanks the Recruitment Program of High-end Foreign
Experts for support.
The work of S.~B.~G.~was supported by the National Natural Science Foundation of China (Grant No.~11675223).

\end{document}